\documentclass[twocolumn]{article}
\usepackage[latin9]{inputenc}
\usepackage{amsmath}
\usepackage{amssymb}
\usepackage{graphicx}
\usepackage{esint}
\makeatletter
\providecommand{\tabularnewline}{\\}

\@ifundefined{date}{}{\date{}}

\usepackage{color}
\usepackage{latexsym}
\usepackage{cite}

\usepackage[left,modulo]{lineno}

\usepackage{color}

\evensidemargin\oddsidemargin \hoffset-22.4mm \voffset-28.4mm
\makeatother

\begin{document}

\title{Propagation of acoustic waves in two waveguides coupled by perforations.
II. Application to periodic lattices of finite length}

\author{Marc Pachebat, Jean Kergomard, \\
 LMA, CNRS, UPR 7051, Aix-Marseille Univ, Centrale Marseille,\\
 13453 Marseille Cedex 13, France. pachebat@lma.cnrs-mrs.fr}
\maketitle
\begin{abstract}
The paper deals with the generic problem  of two waveguides
coupled by perforations, which can be perforated tube mufflers without
or with partitions, possibly with absorbing materials. Other examples are ducts with branched resonators of honeycomb cavities, which can be coupled or not, and splitter silencers. Assuming low frequencies, only one mode is
considered in each guide. The propagation in the two waveguides can be
very different, thanks e.g. to the presence of constrictions. The model is a discrete, periodic one, based upon 4th-order impedance matrices and their diagonalization. All the calculation is analytical, thanks to the partition of the matrices in 2nd-order matrices, and allows the treatment of a very wide types of problems. 
Several aspects are investigated: the local or non-local character of the reaction of one guide to the other; the definition of a coupling coefficient;  the effect of finite size when a lattice with $n$ cells in inserted into an infinite guide; the relationship between the Insertion Loss and the dispersion.  The assumptions are as follows: linear acoustics, no mean flow, rigid wall. However  the effect of the series impedance of the perforations, which is generally ignored, is taken into account, and is discussed. When there are no losses,  it is shown that, for symmetry reasons,  the cutoff frequencies depend on either the series impedance or the shunt admittance, and are the eigenfrequencies of the cells of the lattice, with zero-pressure or zero-velocity at the ends of the cells. 
\end{abstract}

\section{Introduction}

The present paper describes an attempt of a generic study of several
problems that are now classical: perforated tube mufflers without
\cite{Sullivan1978,Sullivan1979,hasegawa1981,Jayaraman1981,Auregan2003}
or with partitions \cite{cheng2015}. They can be with absorbing materials
\cite{Kirby2009,Ji2010,Jiang2010}. Related problems are tubes with
branched resonators, which can be uncoupled \cite{Fang2006,Wang2014}
or coupled \cite{Griffin2000}, or with honeycomb cavities \cite{TaggFaulkner1981,Jones}.
Other kind of systems are splitter silencers with perforated facing
\cite{Ko1975,Auregan2001,Kim2006,Kirby2005a,Kirby2014,Binois2015}.
This generic problem is that of a periodic lattice of two waveguides
coupled by perforations. 

\begin{figure}[t]
\begin{centering}
\includegraphics[width=5.1cm]{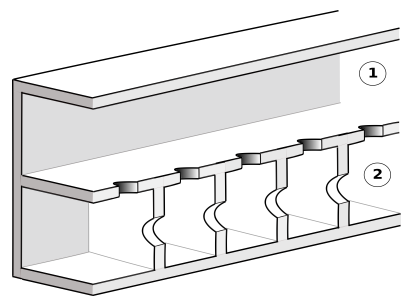} 
\par\end{centering}

\caption{Sectional view of the type of lattice under study : lateral perforations
(dark grey) couple two waveguides periodically along the direction
of propagation. Each waveguide have its own acoustic properties. In
Guide 2, diaphragms are periodically spaced along the direction of
propagation\label{fig:Sectional-view-of}}
\end{figure}

Assuming low frequencies, only one mode is
considered in each guide, therefore the system in study is a system
with two coupled modes. The propagation in the two waveguides can be
very different, thanks to the presence of constrictions, diaphragms, porous material,
partitions or other type of obstacles (see Figure \ref{fig:Sectional-view-of}).
Following Sullivan \cite{Sullivan1979}, we use a discrete, periodic
model based on 4th-order transfer matrices. However the product of
these transfer matrices can lead to diverging products when evanescent
modes are present, and this can be avoided by combining a decoupling approach, i.e., a diagonalization, and then the transformation of a transfer
matrix into an impedance matrix for the finite-length lattice. The
decoupling approach was used also in a continuous modeling (\cite{Peat1988,Jayaraman1981},
see also \cite{Pierce1991} p 356).

The papers aims at showing that it is possible to use an analytical formulation for a very wide class of problems, with the illustration of basic examples of coupled waveguides.
Thanks to  a discrete model, the diagonalization
of 4th-order transfer matrices can be found analytically, by using the partition of these matrices into 2nd-order
matrices. The  study of  the coupling between two guides especially 
involves an analysis of the local or non-local character
of the coupling. Generally speaking, coupling is obviously strong
when the perforations are wide, but also when propagation in the two
guides is rather similar (i.e., the two propagation constants are close). This analysis is done for lattices of finite length, focusing on the behaviour relationships between the Insertion Loss coefficient and the dispersion curves and frequency bands.

A major difficulty is the modeling of the perforations. Semi-empirical
formulas are generally used \cite{Sullivan1979,Maa1998,Dickey2001,Peat2006,Denia2007,bravo2016},
especially when there is a mean flow. One focus of the present paper
is on the role of the series impedance of the perforation \cite{Kergomard1994},
which can be ignored in a continuous model, but not in a discrete
model. A priori this impedance, due to the anti-symmetric field in
the perforation, must be accounted for the  case of wide and well spaced
perforations. To our knowledge, no paper used the complete model found
in the paper published in 1994 \cite{Kergomard1994}. However for
a similar problem in musical acoustics, the effect of the series impedance
of tone-holes of woodwind instruments, can be significant \cite{dubos1999,nederveen1998,debut2005}.

The values of the perforation shunt admittance and series impedance
are not discussed in detail in this paper, but the values given in
\cite{Kergomard1994} are sufficient for a discussion (exact values
were given for the 2D, rectangular case at low frequencies, but for
the cylindrical case, only approximate formulas were proposed).  

Several papers are concerned with more general systems with more than
two guides, or with 2D silencers, in particular for applications to
metamaterials \cite{Liu2000,Wu2002,Kar2005,Fang2006,Wu2010,Chen2010,lafarge2013,groby2015}.
They are not discussed here. Concerning a general view on 1D periodic
structures, we refer to classical references \cite{Brillouin1953a,Elachi1976}.

The assumptions are as follows: linear acoustics, no mean flow, rigid
walls. However the diagonalization is done in a very wide, linear
framework. The basic geometry and the model used are described in
Section \ref{sec:Generic-geometry;-model}, with the definition of
the transfer matrix of a lattice cell. Section \ref{sec:Infinite-periodic-lattice}
derives the eigenvalues and eigenvectors of a cell, using the more
general result given in Appendix \ref{sec:Appendix:-Derivation-of}. For the case of lossless guides, the cut-off frequencies are determined.

For a finite lattice of $n$ cells,  the impedance matrix is derived by using the calculation of the transfer matrix calculated given in Appendix 
\ref{sec:Appendix:-Transfer-matrix}. Finally the insertion loss of the lattice into an infinite waveguide is derived.
Section \ref{III} proposes a theoretical analysis of the coupling
between the two guides, focusing on the effect of the series
impedance; on a definition of a coupling coefficient; and on a derivation
of a condition for a local reaction. Finally numerical simulations of application examples are presented in Section \ref{sec:Applications}, with an analysis of the insertion loss with respect
to the nature of the two modes.

\section{Generic geometry; model and notations\label{sec:Generic-geometry;-model}}

\subsection{Geometry}

\begin{figure}[t]
\begin{centering}
\includegraphics[width=7.5cm]{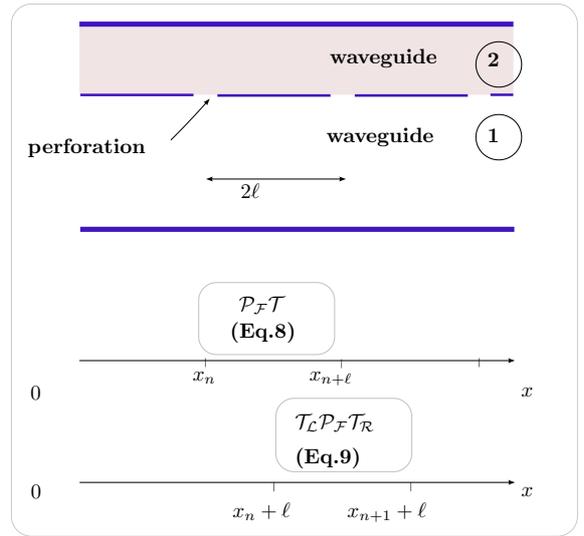} 
\par\end{centering}
\caption{Basic geometry; an asymmetric cell includes one perforation followed
by a length $2\ell$ of tube, with the transfer matrix $P\mathcal{T}$
between abscissas $x_{n}$and $x_{n+1}$, while a symmetric cell includes
one perforation between two lengths $\ell$ of tube, with the transfer
matrix $T_{L}\mathcal{P}_{\mathcal{F}}\mathcal{T_{R}}$, between abscissas
$x_{n}+\ell$ and $x_{n+1}+\ell$. \label{fig:Transfer-Matrix-symmetrical-and--1}}
\end{figure}

The two guides are coupled by perforation, as shown in Fig. \ref{fig:Transfer-Matrix-symmetrical-and--1}.
When their cross section is uniform, the waves are planar at an axial
distance from perforations larger that the transverse dimensions, so
that the evanescent modes due to perforations vanish. When the cross
section is not uniform, the change in cross-section area needs to
be sufficiently far from the perforation, i.e., at an axial distance
larger than the transverse dimension. The propagation in the guides
is characterized by the effective density $\rho_{i}$ and the speed
of sound $c_{i}$ (the subscript $i$ $=1,2$). The change in cross
section allows various situations to be created, as shown in Figure
\ref{fig:Three-Cases}. When the propagation is identical is the two
guides, the lattice is homogeneous, while when diaphragms are present
in one guide only, the lattice is non-homogeneous. The case of branched
resonators without longitudinal coupling between them is a limit case,
with a local reaction of Guide 2 on Guide 1.

\subsection{Model for a perforation}

The general model, valid in harmonic regime, is developed in \cite{Kergomard1994}.
It is summarized hereafter, with similar notations. Four basic quantities
are chosen to be the coefficients $p$ and $v$ of the planar mode
for the acoustic pressure and velocity, respectively, in the two guides.
They build a 4th-order vector, $\mathcal{V}$, as follows: 
\begin{equation}
\ \mathcal{V}=\left(\begin{array}{c}
\mathbf{V}_{1}\\
\mathbf{V}_{2}
\end{array}\right)\text{ \ where }\mathbf{V}_{i}=\left(\begin{array}{c}
p_{i}\\
v_{i}
\end{array}\right),\text{ }\label{eq:E1}
\end{equation}
$i=1,2.$ The following notations are chosen: calligraphic characters
correspond to 4th-order matrices and vectors, while bold characters
correspond to 2nd-order matrices and vectors (e.g. $\mathcal{I}$
and \textbf{I} are the identity matrices of order 4 and 2, respectively);
other quantities are scalar. For a periodic medium made of asymmetric
lattice cells, one perforation at $x_{n}$ is followed by a portion
of length 2$\ell$ of separated waveguides (see Fig. \ref{fig:Transfer-Matrix-symmetrical-and--1}).
The vectors $\mathcal{V}$ are related by 4th-order matrices.

For a perforation the following relationship is derived in \cite{Kergomard1994}
(the subscripts L and R correspond to the left side and right side
 of a perforation, respectively): 
\begin{eqnarray}
\mathcal{V}_{L} & = & \mathcal{P}_{\mathcal{F}}\text{ \ }\mathcal{V}_{R}\label{eq:E2}\\
\textrm{where} &  & \text{ }\mathcal{P}_{\mathcal{F}}=\left(\begin{array}{cc}
(\gamma_{1}+\gamma_{2}\mathbf{M)} & \gamma_{2}(\mathbf{I}-\mathbf{M)}\\
\gamma_{1}(\mathbf{I}-\mathbf{M)} & (\gamma_{2}+\gamma_{1}\mathbf{M)}
\end{array}\right),\\
\text{with }\gamma_{1,2} & = & \frac{S_{1,2}}{S_{1}+S_{2}},\text{ }\label{eq:E3}\\
\mathbf{M} & = & \mathbf{I+}\frac{2Z_{a}Y_{s}}{1-Z_{a}Y_{s}}\mathbf{K},\:\text{ }\mathbf{K}\mathbf{=}\left(\begin{array}{cc}
1 & Y_{s}^{-1}\\
Z_{a}^{-1} & 1
\end{array}\right).
\end{eqnarray}
$S_{1,2}$ are the cross-section areas of the guides. This model considers
the effect of a perforation as localized at the abscissa of the perforation
center, as explained in \cite{Schwinger1968}. $Z_{a\text{ }}$and
$Y_{s}$ are the series impedance and shunt admittance of the perforation,
respectively (these quantities are \emph{specific} impedance and \textit{specific
}admittance, i.e., ratios pressure/velocity and velocity/pressure,
respectively). Both are acoustic masses and correspond to the anti-symmetric
and symmetric pressure field in the perforation, respectively. When
$Z_{a}=0$, $Y_{s}$ produces a jump in velocity inside each guide,
from the left to the right of the perforation, along the guide axis.
In a dual way, when $Y_{s}=0,$ $Z_{a}$ produces a jump in pressure
inside each guide, from the left to the right of the perforation. Reciprocity is assumed, therefore the determinant of $\mathcal{P}_{\mathcal{F}}$ is unity.

\subsection{Model for the propagation in the waveguides}

For a non-perforated portion of the waveguides, between abscissas
$x_{n}$ and $x_{n+1}$, the following 4th-order matrix relationship
is written as: 
\begin{equation}
\mathcal{V}_{R,n}=\mathcal{TV}_{L,n+1}\ \text{where}\ \mathcal{T=}\left(\begin{array}{cc}
\mathbf{T}_{1} & \mathbf{0}\\
\mathbf{0} & \mathbf{T}_{2}
\end{array}\right).\label{eq:E5}
\end{equation}
The general transfer matrices 
\begin{equation}
\mathbf{T}_{1,2}=\left(\begin{array}{cc}
A_{1,2} & B_{1,2}\\
C_{1,2} & D_{1,2}
\end{array}\right)\label{eq:E6}
\end{equation}
($i=1,2)$ are of 2nd-order and describe the propagation within Guides
1 and 2. The coefficients $B_{1,2}$ are specific impedances, while
the coefficients $C_{1,2}$ are specific admittances. In the separated
portion, the geometry may be various, e.g., may includes discontinuities
and/or dissipation.

\%end{center}

\begin{figure}[t]
\begin{centering}
\includegraphics[width=7.5cm]{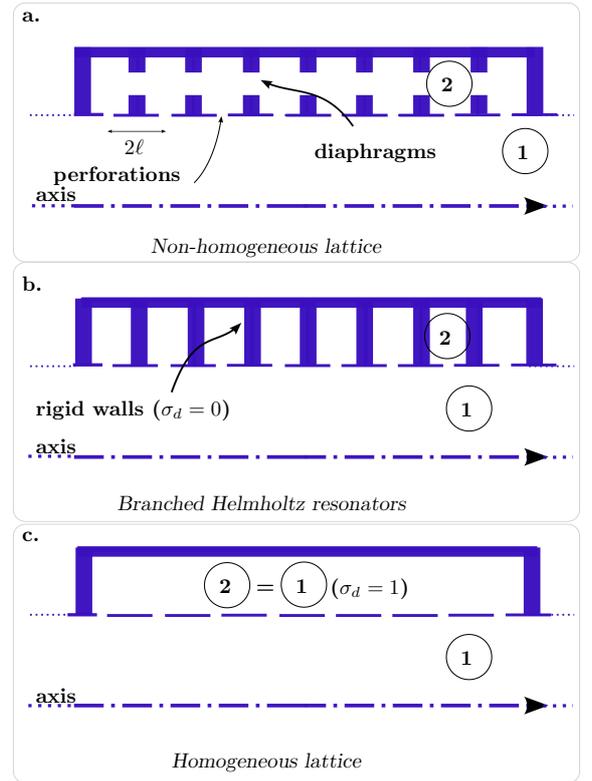} 
\par\end{centering}
\caption{(a) Three simple cases of coupled waveguides. Non-homogeneous lattice, (b) branched Helmholtz resonators, and
(c) homogeneous lattice. Case (b) is a limit case of perfectly local reaction. \label{fig:Three-Cases}}
\end{figure}

\subsection{Model for the propagation with perforations: asymmetric and symmetric
cells \label{asc}}

For a periodic medium, two types of cells can be considered (see Fig.
\ref{fig:Transfer-Matrix-symmetrical-and--1}): i) an asymmetric cell,
involving a perforation followed by a portion of tubes of length $2\ell$;
ii) a symmetric cell involving a portion of tubes of length $\ell$,
then perforation, then a portion of tube of length $\ell$.
The complete transfer matrix of an asymmetric cell (see Fig.\ref{fig:Transfer-Matrix-symmetrical-and--1})
can be characterized by the equation: 
\begin{equation}
\mathcal{V}_{L,n}=\mathcal{P}_{\mathcal{F}}\mathcal{TV}_{L,n+1}\;.\label{eq:E7}
\end{equation}
The case of a symmetric cell is more particular, but remains very
general. It will be used for the diagonalization (see next section).
For such a cell, between abscissas $x_{n}+\ell$ and $x_{n+1}+\ell$,
the transfer matrix relationship is given by: 
\begin{equation}
\mathcal{V}_{n-1}=\mathcal{\mathrm{(}T_{L}}\mathcal{P}_{\mathcal{F}}\mathcal{T_{R}})\mathcal{V}_{n}\,,\label{eq:E26}
\end{equation}
where $\mathcal{V}_{n}=^{t}(p_{1n},v_{1n},p_{2n},v_{2n})$, defined in Equation (\ref{eq:E1}), is considered at
mid-distance (abscissa $x_{n}+\ell$) between two perforations. The
transfer matrix $\mathcal{T}$ \ (Equation (\ref{eq:E5})) describing
the uncoupled propagation over distance $2\ell$ between two neighboring
perforations is therefore the product of the two transfer matrices:
\begin{equation}
\mathcal{T}=\mathcal{T_{R}}\mathcal{T_{L}},\label{eq:E27}
\end{equation}
where $\mathcal{T_{L}}$ (resp.$\mathcal{T_{R}}$) describes the uncoupled
propagation over the distance $\ell$ situated on the left (resp.
right) of one perforation. Since $\mathcal{T}$ \ is block-diagonal
(Equation (\ref{eq:E5})), we can adopt the same decomposition for
2nd-order blocks, namely $\mathbf{T}_{1}=\mathbf{T}_{R1}\mathbf{T}_{L1}$,
and $\mathbf{T}_{2}=\mathbf{T}_{R2}\mathbf{T}_{L2}$. Moreover, in
order for the  cell to be symmetric, we generalize the concept of reversed
four-terminal explained in \cite{Brillouin1953a}. The matrices $\mathbf{T}_{L}$
($i=1,2)$ need to be proportional to the invert of the matrices $\mathbf{T}_{R}$,
with a change in sign for the x-axis, and with the same determinant
$\delta$. This means: 
\begin{equation}
\left(\begin{array}{cc}
A_{R} & B_{R}\\
C_{R} & D_{R}
\end{array}\right)=\left(\begin{array}{cc}
D_{L} & B_{L}\\
C_{L} & A_{L}
\end{array}\right),\label{JK9}
\end{equation}
With this condition the matrix $\mathbf{T}$ is symmetric: 
\begin{equation}
A=D=A_{L}D_{L}+B_{L}C_{L};\:B=2B_{L}D_{L}\:C=2C_{L}A_{L}.\label{JK99}
\end{equation}

Moreover reciprocity is assumed (in particular no flow is present), i.e., the determinant is
unity, as well as the determinant of the 4th-order matrix $\mathcal{T}$. More general cases are investigated in Appendix \ref{sec:Appendix:-Derivation-of}.

\section{Infinite periodic lattice: eigenvalues and eigenvectors\label{sec:Infinite-periodic-lattice}}

\subsection{Eigenvalues and eigenvectors}

In this section, we are searching for the diagonal form of the transfer
matrix $\mathcal{T}_{L}\mathcal{P}_{\mathcal{F}}\mathcal{T_{R}}$
(Equation (\ref{eq:E26})) for an elementary, symmetric cell (with reciprocity)
of the periodic lattice shown on Fig.(\ref{fig:Transfer-Matrix-symmetrical-and--1}):
\begin{equation}
\mathcal{T_{L}}\mathcal{P}_{\mathcal{F}}\mathcal{T_{R}}\overset{def}{=}\mathcal{E}\mathcal{D}\mathcal{E}^{-1}\label{eq:E9}
\end{equation}
with 
\begin{eqnarray}
\mathcal{D} & = & \left(\begin{array}{cccc}
\lambda^{(1)} & 0 & 0 & 0\\
0 & \lambda^{(2)} & 0 & 0\\
0 & 0 & \lambda^{(3)} & 0\\
0 & 0 & 0 & \lambda^{(4)}
\end{array}\right)\,,\text{ and}\label{E10}\\
\mathcal{E} & = & \left(\begin{array}{cccc}
\mathcal{W}^{(1)} & \mathcal{W}^{(2)} & \mathcal{W}^{(3)} & \mathcal{W}^{(4)}\end{array}\right)\,.\label{E11}
\end{eqnarray}
$\lambda^{(i)\text{ }}$ are the eigenvalues and $\mathcal{W}^{(i)}$
($i=1..4$) are the eigenvectors. The detailed calculation is derived
in Appendix \ref{sec:Appendix:-Derivation-of} for the most general
case (no reciprocity is required). Since for the perforation matrix (Equation (\ref{eq:E2})),
$\det\mathcal{P_{F}}=1$, the eigenvalues of the diagonal matrix $\mathcal{D}$
(Equation (\ref{E10})) can be grouped by inverse pairs when reciprocity
holds for the elementary cell $\left(\lambda^{(1)},\lambda^{(2)}\right)=(\lambda,1/\lambda)$
and $\left(\lambda^{(3)},\lambda^{(4)}\right)=(\lambda^{\prime},1/\lambda^{\prime})$
(see \cite{Kergomard1994}). Each pair corresponds to opposite propagation
directions of an eigenmode along the lattice axis. They are
denoted $\lambda=\exp(\Gamma)$ \ and $\lambda^{\prime}=\exp(\Gamma^{\prime})$.
This leads to te following dispersion equation for the unknowns $\Gamma$
and $\varGamma'$:

\begin{equation}
\det\mathbf{T}_{0}=A_{0}D_{0}-B_{0}C_{0}=0,\label{JK4-1}
\end{equation}
where
\begin{alignat}{1}
A_{0} & =D_{0}=-\sinh\Gamma\,\left[\gamma_{2}/Q_{1}+\gamma_{1}/Q_{2}\right]\notag\\
B_{0} & =\frac{1}{Y_{s}}-\gamma_{2}B_{1}/Q_{1}-\gamma_{1}B_{2}/Q_{2}\label{eq:E33}\\
C_{0} & =\frac{1}{Z_{a}}-\gamma_{2}C_{1}/Q_{1}-\gamma_{1}C_{2}/Q_{2}\,\notag
\end{alignat}
 with $Q_{i}=\textrm{(cosh}\Gamma-A_{i})$. The eigenvector matrix
is found to be: 
\begin{equation}
\mathcal{E=}v_{0}\left(\begin{array}{cccc}
z_{1} & z_{1} & z_{1}^{\prime} & z_{1}^{\prime}\\
h_{1} & -h_{1} & h_{1}^{\prime} & -h_{1}^{\prime}\\
-z_{2} & -z_{2} & -z_{2}^{\prime} & -z_{2}^{\prime}\\
-h_{2} & +h_{2} & -h_{2}^{\prime} & +h_{2}^{\prime}
\end{array}\right)\widetilde{\mathcal{D}}^{-1/2}\;\textrm{,}\label{356}
\end{equation}

\begin{eqnarray}
z_{1} & = & \frac{1}{\gamma_{1}Q_{1}}\left[B_{R1}\cosh(\Gamma/2)-w_{0}D_{R1}(\sinh\Gamma/2)\right]\,,\label{JK98}\\
h_{1} & = & \frac{1}{\gamma_{1}Q_{1}}\left[A_{R1}(\sinh\Gamma/2)-w_{0}C_{R1}\cosh(\Gamma/2)\right]\,\label{JK97}\\
w_{0} & = & B_{0}/A_{0}=D_{0}/C_{0}.
\end{eqnarray}
Similar expressions can be found for $h_{2}$ and $z_{2}$. For $h'_{1}$
and $z'_{1}$, $\Gamma$ is changed in $\Gamma'$ and $w_{0}$ in
$w'_{0}$, and similarly for the quantities with subscript 2. The
matrix $\widetilde{\mathcal{D}}^{-1/2}$ corresponds to a shift of
an eigenvector by one half-cell. $v_{0}$ is an arbitrary constant
with the dimension of a velocity. Notice that because the eigenvectors
are defined apart from a multiplicative constant, three quantities
define an eigenvector. Coming back to the definition of the physical-quantity
vectors (see Equation (\ref{eq:E1})), we deduce the following interpretations: 
\begin{itemize}
\item The ratio $z_{1}/h_{1}$ is the (specific) characteristic impedance
in Guide 1 for the first propagation constant $\Gamma$; 
\item Because the second eigenvalue corresponds to a change in sign of the
propagation constant $\Gamma$, the corresponding characteristic impedance
is $-z_{1}/h_{1}$, as expected; 
\item Similar remarks hold for subscript 2 and superscript '; 
\item With the two characteristic impedances, the last quantity defining
an eigenvector is the velocity ratio $-h_{1}/h_{2};$ this ratio is
identical for the two waves with opposite propagation constants. 
\end{itemize}
In order to calculate the constant $\Gamma$, Equation (\ref{JK4-1})
can be re-written as a 2nd-order equation for the unknown $\cosh(\Gamma)$.
For this purpose the terms proportional to $\gamma_{1}^{2}$ and $\gamma_{2}^{2}$
can be rearranged by using the relations $\gamma_{1}+\gamma_{2}=1$
and $\sinh^{2}\Gamma-B_{i}C_{i}=\cosh^{2}\Gamma-A_{i}^{2}$. The following
equation is obtained: 
\begin{alignat}{1}
\left(1-Y_{s}Z_{a}\right)\cosh^{2}\Gamma+\nonumber \\
-\left[A_{1}+A_{2}+\gamma_{2}E_{1}+\gamma_{1}E_{2}+d_{12}\right]\cosh\Gamma\nonumber \\
+\gamma_{1}\gamma_{2}Y_{s}Z_{a}\left[(B_{1}C_{2}+B_{2}C_{1})+2+2A_{1}A_{2}\right]\nonumber \\
+A_{1}A_{2}\left(1+Y_{s}Z_{a}\right)+\gamma_{2}A_{2}E_{1}+\gamma_{1}A_{1}E_{2} & =0,\label{eq:JKP}
\end{alignat}
where $d_{12}=(A_{1}-A_{2})(\gamma_{2}-\gamma_{1})Y_{s}Z_{a}$ and
$E_{i}=Y_{s}B_{i}+Z_{a}C_{i}$. Thanks to Equation (\ref{eq:JKP}),
general solutions $\cosh\Gamma$ and $\cosh\Gamma'$, for the two
modes $\Gamma$ and $\Gamma'$ of the lattice can be written explicitly.

\subsection{Reciprocity relationships \label{reci}}

Reciprocity is related to the choice of matrices $\mathbf{T}_{1}$,
$\mathbf{T}_{2}$ and $\mathbf{M}$, with a determinant equal to unity.
In order to find the consequences on the eigenvectors of a cell, we
start from the classical reciprocity equation valid for guides without
flow. We write it on the surface $\Sigma$ of a cell (e.g., a symmetric
cell): 
\begin{equation}
\int\int_{\Sigma}\left(p^{(i)}\overrightarrow{v^{(j)}}-p^{(j)}\overrightarrow{v^{(i)}}\right)d\overrightarrow{\Sigma}=0.\label{500}
\end{equation}
The superscripts $i$ and $j$ correspond to two different situations.
For instance two situations where only one eigenmode exists can be
chosen. The integral vanishes on all rigid walls, therefore it is
limited to the input and output of a cell. The term in parenthesis
in Equation (\ref{500}) is the same for the output surface and the
input surface, apart from the factor $-\exp(\Gamma^{(i)})\exp(\Gamma^{(j)}).$
Therefore it is possible to factorize the term $\left[1-\exp(\Gamma^{(i)})\exp(\Gamma^{(j)})\right]$,
and for the eigenmodes corresponding to $\Gamma$ and $-\Gamma$,
Equation (\ref{500}) is trivial because this term vanishes. It remains
to solve the following equation: 
\begin{equation}
\int\int_{S_{1}+S_{2}}\left(p\overrightarrow{v^{\prime}}-p^{\prime}\overrightarrow{v}\right)d\overrightarrow{\Sigma}=0\label{502}
\end{equation}
for the modes corresponding to $\Gamma$ and $\Gamma^{\prime}$, and
to $\Gamma$ and $-\Gamma^{\prime}$.
 Using the expressions (\ref{356})
of the eigenvectors, the following equations are obtained: 
\begin{equation}
\gamma_{1}z_{1}h'_{1}=-\gamma_{2}z_{2}h_{2}^{\prime}\text{ ; }\gamma_{1}z'_{1}h_{1}=-\gamma_{2}z_{2}^{\prime}h_{2}.\label{501}
\end{equation}
A direct checking of these equations is heavy. Reciprocity implies
also the symmetry of the impedance matrix, as shown in Section \ref{sub:Impedance-matrix-for}.

\subsection{Lossless lattices; cut-off frequencies\label{dtw}}

Up to now the considered lattice is can be lossy, when one of the
coefficients defining a cell is complex. For lossless waveguides,
several types of waves can exist \cite{Kergomard1994}. When reciprocity
holds, each of the two modes with propagation constant $\Gamma$ and
$\Gamma^{\prime}$ can be either propagating or evanescent. In the
case of two evanescent waves the possibility for the propagation constant
to be complex was found: the energy flux in each guide decreases exponentially,
but is not zero (its sign is opposite in the two guides, ensuring
the energy conservation).

Ref. \cite{Kergomard1994} studied the particular case of an homogenous
lattice, i.e., a lattice with identical transfer matrices $\mathbf{T}$
in the two guides. This happens for example when the guides are straight
guides with the same sound speed and density. In this case, there
is at least one propagating wave, and the decomposition of the propagation
into two modes (one is planar, the other one is called the ``flute'' mode)
is valid even for a lattice with irregular perforations.

The cut-off frequencies are given by $\textrm{cosh}\Gamma=\pm1$,
i.e., $\Gamma=0$ or $\Gamma=j\pi$. Therefore, according to Equations
  (\ref{eq:E33}), $A_{0}=D_{0}=0$, and the dispersion equation (\ref{JK4-1})
implies either $B_{0}=0$ or $C_{0}=0.$ Writing $\cosh\Gamma=\pm1$
in Equation (\ref{eq:E33}), and using Equations   (\ref{JK9},\ref{JK99})
with the property $\det(\mathbf{T}_{L})=1$, the cut-off frequencies
are given by one of the four following equations: 
\begin{align}
\Gamma=0;\:B_{0}=0;\Rightarrow\frac{1}{Y_{s}}+\gamma_{2}\frac{D_{L1}}{C_{L1}}+\gamma_{1}\frac{D_{L2}}{C_{L2}} & =0;\label{JK26}\\
\Gamma=0;\:C_{0}=0;\Rightarrow\frac{1}{Z_{a}}+\gamma_{2}\frac{A_{L1}}{B_{L1}}+\gamma_{1}\frac{A_{L2}}{B_{L2}} & =0;\label{JK27}\\
\Gamma=j\pi;\:B_{0}=0;\Rightarrow\frac{1}{Y_{s}}+\gamma_{2}\frac{B_{L1}}{A_{L1}}+\gamma_{1}\frac{B_{L2}}{A_{L2}} & =0;\label{JK28}\\
\text{\ensuremath{\Gamma}=j\ensuremath{\pi};\:\ensuremath{C_{0}}=0;\ensuremath{\Rightarrow}}\frac{1}{Z_{a}}+\gamma_{2}\frac{C_{L1}}{D{}_{L1}}+\gamma_{1}\frac{C_{L2}}{D_{L2}} & =0.\label{JK29}
\end{align}
The characteristic impedances of the two guides, $z_{i}/h_{i}$
(see Equations.(\ref{JK98},\ref{JK97}), are found to be either infinite
ou zero. It is interesting to interpret these results. Consider the
example of Equation (\ref{JK26}). Because $\Gamma=0$, for an infinite
lattice, $p_{n}=p_{n+1}$ in each guide, and because the characteristic
impedance is infinite, the velocity vanishes at the extremities of
the cell. Consequently, if there is a opening at the extremity of
the cell, this cut-off does not depend on the opening. It can be checked
that this equation gives the eigenfrequency of the cell when it is
closed at their extremities (infinite impedance). The second and the
third terms of Equation (\ref{JK26}) correspond to the impedance in Guide 1 and 2, respectively,
at the abscissa of the perforation, calculated by projecting the infinite
impedance at the end of the cell to the perforation abscissa. Moreover the pressure field in the
cell being symmetrical, the series impedance $Z_{a}$ does not intervene. 

Similar interpretation can be done for the three other equations,
using the duality pressure/velocity.

\section{Impedance matrix of a lattice on $n$ cells; insertion into an infinite
waveguide\label{sub:Transfer-matrix-for}}

\subsection{Impedance matrix\label{sub:Impedance-matrix-for}}

In order to derive the (\textit{acoustic}) impedance matrix of a lattice
of $n$ cells, the vector $\mathcal{V}$ (Equation (\ref{eq:E1}))
is replaced by a vector $\widetilde{\mathcal{V}}$ defined as follows:
\begin{equation}
\ \widetilde{\mathcal{V}}=\left(\begin{array}{c}
\mathbf{P}\\
\mathbf{U}
\end{array}\right)\text{ \ where }\mathbf{P}=\left(\begin{array}{c}
p_{1}\\
p_{2}
\end{array}\right)\text{ and }\mathbf{U}=\left(\begin{array}{c}
u_{1}\\
u_{2}
\end{array}\right),
\end{equation}
where $u_{i}=S_{i}v_{i}$ ($i=1,2)$ are the flow rates. In Appendix
\ref{sec:Appendix:-Transfer-matrix} it is shown that for these vectors
the transfer matrix relationship can be written as:

\begin{equation}
\widetilde{\mathcal{V}}_{0}=\left(\begin{array}{cc}
\mathbf{Z} & \mathbf{0}\\
\mathbf{0} & \mathbf{G}
\end{array}\right)\left(\begin{array}{cc}
\mathbf{C_{n}} & \mathbf{S_{n}}\\
\mathbf{S_{n}} & \mathbf{C_{n}}
\end{array}\right)\left(\begin{array}{cc}
\mathbf{Z} & \mathbf{0}\\
\mathbf{0} & \mathbf{G}
\end{array}\right)^{-1}\widetilde{\mathcal{V}}_{n}\,,\label{450-1}
\end{equation}

with 
\begin{align}
\mathbf{C}_{n} & =\left(\begin{array}{cc}
\cosh n\Gamma & 0\\
0 & \cosh n\Gamma^{\prime}
\end{array}\right),\label{600}
\end{align}
 
\begin{align}
\mathbf{S}_{n} & =\left(\begin{array}{cc}
\textrm{sinh}n\Gamma & 0\\
0 & \textrm{sinh}n\Gamma^{\prime}
\end{array}\right),\label{601}
\end{align}

\begin{align}
\mathbf{Z} & =\left(\begin{array}{cc}
z_{1} & z_{1}^{\prime}\\
-z_{2} & -z_{2}^{\prime}
\end{array}\right),\mathbf{\:G}=\left(\begin{array}{cc}
g_{1} & g_{1}^{\prime}\\
-g_{2} & -g_{2}^{\prime}
\end{array}\right),\label{602}
\end{align}
 if $g_{i}=S_{i}h_{i}$. This \textit{acoustic} impedance matrix is directly derived from this
transfer matrix. It is chosen for two reasons: i) the impedance matrix
avoids numerical difficulties that appear using transfer matrix products,
for strongly evanescent eigenmodes and a large number of cells; ii)
the impedance matrix makes easy  the boundary conditions to be introduced 
at each end of the lattice. Consider two 4th-order vectors$\,{}^{t}\left(\begin{array}{c}
\mathbf{P_{0}}\end{array}\mathbf{U}_{0}\right)$ and $\,{}^{t}\left(\begin{array}{c}
\mathbf{P_{n}}\end{array}\mathbf{U}_{n}\right)$ related by a (general) matrix as follows: 
\begin{equation}
\left(\begin{array}{c}
\mathbf{P_{0}}\\
\mathbf{U_{0}}
\end{array}\right)=\left(\begin{array}{cc}
\mathbf{A} & \mathbf{B}\\
\mathbf{C} & \mathbf{D}
\end{array}\right)\left(\begin{array}{c}
\mathbf{P_{n}}\\
\mathbf{U}_{n}
\end{array}\right)\,,\label{451}
\end{equation}
where $\mathbf{A}$, $\mathbf{B}$, $\mathbf{C}$ and $\mathbf{D}$
are 2nd order-matrices. This expression is equivalent to:

\begin{equation}
\left(\begin{array}{c}
\mathbf{P_{0}}\\
\mathbf{P}_{n}
\end{array}\right)=\left(\begin{array}{cc}
\mathbf{A}\mathbf{C}^{-1} & \left[\mathbf{B}-\mathbf{A}\mathbf{C}^{-1}\mathbf{D}\right]\\
\mathbf{C^{-1}} & \mathbf{-C}^{-1}\mathbf{D}
\end{array}\right)\left(\begin{array}{c}
\mathbf{U_{0}}\\
\mathbf{U}_{n}
\end{array}\right)\,.\label{eq:E52}
\end{equation}
Applying this result to the transfer matrix (Equation (\ref{450-1})),
the impedance matrix of the lattice of $n$ cells is obtained:

\begin{multline}
\left(\begin{array}{c}
p_{1,0}\\
p_{2,0}\\
p_{1,n}\\
p_{2,n}
\end{array}\right)=\mathcal{Z}\left(\begin{array}{cc}
\mathbf{C}_{n}\mathbf{S}_{n}^{-1} & \mathbf{-S}_{n}^{-1}\\
\mathbf{S}_{n}^{-1} & -\mathbf{S}_{n}^{-1}\mathbf{C}_{n}
\end{array}\right)\mathcal{G}^{-1}\left(\begin{array}{c}
u_{1,0}\\
u_{2,0}\\
u_{1,n}\\
u_{2,n}
\end{array}\right)\\
\mathcal{\textrm{where }Z=\left(\begin{array}{cc}
\mathbf{Z} & \mathbf{0}\\
\mathbf{0} & \mathbf{Z}
\end{array}\right)}\textrm{ and }\mathcal{G=\left(\begin{array}{cc}
\mathbf{G} & \mathbf{0}\\
\mathbf{0} & \mathbf{G}
\end{array}\right)},\label{eq:E53}
\end{multline}
where the identity $\left[\mathbf{S}_{\mathrm{n}}-\mathbf{C}_{\mathrm{n}}\mathbf{S}_{\mathrm{n}}^{\mathrm{-1}}\mathbf{C}_{\mathrm{n}}\right]=\mathbf{-S}_{\mathrm{n}}^{\mathrm{-1}}$
is used. Actually, because of the different sign before $\mathbf{S}_{n}^{-1}$
in the second diagonal, this impedance matrix is anti-symmetric (with
a change in the orientation of the velocities at the extremity $n$,
the impedance matrix would become symmetric). Notice that the matrix
$\mathcal{Z}$ has the dimension of a specific impedance,
while the matrix $\mathcal{G}$ has the dimension of the inverse of
an area. Furthermore, for evanescent modes (real $\Gamma$)  the  ratios $\textrm{cosh}\Gamma/\textrm{sinh}\Gamma$ and $1/\textrm{sinh}\Gamma$
do no diverge when $n$ tends to infinity, unlike the coefficients of the transfer matrix. 


\begin{figure}[t]
\begin{centering}
\includegraphics[width=7.5cm]{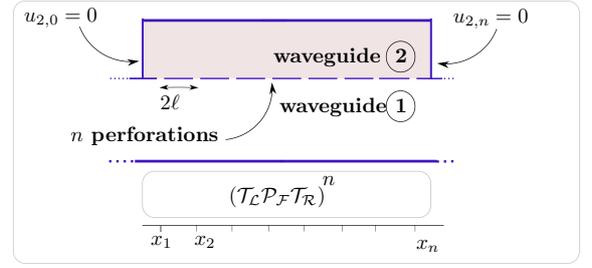} 
\par\end{centering}
\caption{Lattice with $n$ cells, with the transfer matrix $\mathcal{\mathrm{(}T_{L}}\mathcal{P}_{\mathcal{F}}\mathcal{T_{R}})^{n}$
and boundary conditions for Guide 2 (see Equations   (\ref{eq:E54}
and \ref{eq:E57})), inserted into an infinite waveguide.\label{fig:Finite-Length-Lattice}}
\end{figure}

\subsection{Lattice of finite length inserted into an infinite waveguide\label{sec:Finite-length-lattice}}

We consider the geometry shown in Figure \ref{fig:Finite-Length-Lattice}.
A lattice of finite length, with $n$ cells, is inserted into an infinite
waveguide so as to act as an acoustic wall treatment. By closing Guide
2 at each end of the lattice (Equation (\ref{eq:E54})) by an impedance
condition (Figure \ref{fig:Finite-Length-Lattice}), the 4th-order impedance
matrix (Equation (\ref{eq:E53})) is reduced to a 2nd-order one, and
the Insertion Loss \cite{Pierce1991} of the finite lattice can be
obtained.

Simple boundary conditions are chosen. Guide 2 is  closed at each
end by setting $u_{2,0}=u_{2,n}=0$ into Equation (\ref{eq:E53}).
The 2nd-order impedance matrix of the finite lattice can be derived:

\begin{equation}
\left(\begin{array}{c}
p_{1,0}\\
p_{1,n}
\end{array}\right)=\left(\begin{array}{cc}
Z_{A} & -Z_{B}\\
Z_{B} & -Z_{A}
\end{array}\right)\left(\begin{array}{c}
u_{1,0}\\
u_{1,n}
\end{array}\right),\text{ with}\label{eq:E54}
\end{equation}

\begin{eqnarray}
 &  & \begin{array}{ccc}
Z_{A} & = & \left[\mathbf{Z}\mathbf{C}_{n}\mathbf{S}_{n}^{-1}\mathbf{G}^{-1}\right]_{11}\\
 & = & \left[\hat{Z}\coth n\Gamma+\hat{Z}^{\prime}\coth n\Gamma^{\prime}\right]
\end{array}\label{eq:E55}\\
 &  & \begin{array}{ccc}
Z_{B} & = & \left[\mathbf{Z}\mathbf{S}_{n}^{-1}\mathbf{G}^{-1}\right]_{11}\\
 & = & \left[\hat{Z}/\sinh n\Gamma+\hat{Z}^{\prime}/\sinh n\Gamma^{\prime}\right],
\end{array}\label{eq:E56}
\end{eqnarray}
where the impedances $\hat{Z}$ and $\hat{Z}^{\prime}$ associated
to each mode result from Equation (\ref{602}) as follows: 
\begin{align}
\hat{Z}=z_{1}g_{2}^{\prime}/(\det\mathbf{G)} & \,\textrm{and\,} & \hat{Z}^{\prime}=z_{1}^{\prime}g_{2}/(\det\mathbf{G)}.\label{eq:ZZp}
\end{align}
Recall that these impedances are \textit{acoustic} impedances (ratio
pressure/flow rate). This particular case of lattice is built as the
combination of two four-terminals with their extremities in series,
each four-terminal corresponding to a propagation mode with constant
$\Gamma$ and $\Gamma^{\prime}$. Expression (\ref{eq:E54}) is then
written in form of a transfer matrix:

\begin{eqnarray}
\left(\begin{array}{c}
p{}_{1,0}\\
u_{1,0}
\end{array}\right) & = & \left(\begin{array}{cc}
A_{s} & B_{s}\\
C_{s} & A_{s}
\end{array}\right)\left(\begin{array}{c}
p{}_{1,n}\\
u_{1,n}
\end{array}\right)\notag\\
 & = & \frac{1}{Z_{B}}\left(\begin{array}{cc}
Z_{A} & (Z_{A}^{2}-Z_{B}^{2})\\
1 & Z_{A}
\end{array}\right)\left(\begin{array}{c}
p{}_{1,n}\\
u_{1,n}
\end{array}\right).\label{eq:E57}
\end{eqnarray}
Let us consider an infinite waveguide with characteristic impedance $\bar{z}_{c1}=\rho_{1}c_{1}/S_{1}$.
The outgoing and incoming plane wave have the amplitudes $p_{1}^{+}=(p_{1}+\bar{z}_{c1}u_{1})/2$
and $p_{1}^{-}=(p_{1}-\bar{z}_{c1}u_{1})/2,$ respectively. Once the
finite lattice of Fig.(\ref{fig:Finite-Length-Lattice}) is inserted,
 the transmission coefficient can be written as: 
\begin{eqnarray}
T & = & \frac{p_{1,n}^{+}}{p_{1,0}^{+}}=\frac{p_{1,0}^{-}}{p_{1,n}^{-}}=\frac{2}{2A_{s}+(B_{s}/\bar{z}_{c1}+C_{s}\bar{z}_{c1})}\\
 & = & \frac{2Z_{B}\bar{z}_{c1}}{(Z_{A}+\bar{z}_{c1}-Z_{B})(Z_{A}+\bar{z}_{c1}+Z_{B})}\,.\label{eq:E66}
\end{eqnarray}
The insertion loss is defined as the ratio of the sound power of the
incident plane wave to that of the transmitted one across the finite
lattice with characteristic impedance $\bar{z}_{c1}$ at its both
ends. It is equal to:

\begin{equation}
IL(\omega)=10\log_{10}\left\vert \frac{1}{T}\right\vert ^{2}\,.\label{eq:E69}
\end{equation}
We notice from Expressions (\ref{eq:E66},\ref{eq:E69}) that when
$Z_{B}=0$ (Equation  (\ref{eq:E56})), i.e., when: 
\begin{equation}
\hat{Z}/\sinh n\Gamma+\hat{Z}^{\prime}/\sinh n\Gamma^{\prime}=0\ ,\label{eq:E70}
\end{equation}
the transmission coefficient is zero, and the associated insertion
loss is infinite (in practice it is limited by losses).

\section{Analysis of the coupling effect; local vs non-local reaction \label{III}}

\subsection{Effect of the series impedance on the coupling\label{sub:Effect-of-Za}}

The respective roles of the series impedance $Z_{a}$ and the shunt
admittance $Y_{s}$ can be discussed qualitatively at the zero-frequency
limit. Exact values are known for the 2D, rectangular case. For the cylindrical case,  we use  approximate values for  two guides with radii
$a_{1}$and $a_{2}$ and the same fluid density $\rho$, which exhibit
the dependence on the parameters \cite{Kergomard1994}: 
\begin{equation}
Y_{s}^{-1}\simeq j\omega\frac{\rho}{r_{p}}S_{1};Z_{a}\simeq-j\omega0.57\rho r_{p}^{2}a_{1}/S_{1},
\end{equation}
 where $j=(-1)^{2}$, $\omega$ is the angular frequency. $r_{p}$
is the radius of the perforation, and $a_{2}$ is assumed to be larger
than $a_{1}$. A first observation is that the product $Y_{s}Z_{a}$
is independent of the frequency and is very small, because it is proportional
to $(r_{p}/a_{1})^{3}.$ As a consequence, at a first approximation,
Equation  (\ref{eq:JKP}) can be simplified in: 
\begin{multline}
(\cosh\Gamma-A_{1}-\gamma_{2}E_{1})(\cosh\Gamma-A_{2}-\gamma_{1}E_{2})\\
=\gamma_{1}\gamma_{2}E_{1}E_{2}.\label{JKb}
\end{multline}
The influence of the series impedance can be estimated by considering
the expression of the quantities $E_{i}.$ Considering the low frequency
case, the guides are reduced to lumped elements, $B_{i}=2j\omega\rho\ell$
is a mass and $C_{i}=2j\omega\ell/\rho c^{2}$ is a compliance ($c$
is the sound speed). It turns out that $Y_{s}B_{i}$ is a ratio of
two masses, while $Z_{a}C_{i}$ is proportional to $\omega^{2}$:
therefore the effect of the series impedance $Z_{a}$ can be neglected
at low frequency. This justifies the following analysis of the coupling
of the two guides with $Z_{a}=0.$ This approximation will be done
from here until to the end of the paper.

\subsection{Eigenvalues and eigenvectors for the simplified model}

What are the conditions for reducing the number of guided modes form
2 to 1? 

If $Z_{a}=0$, the perforation matrix (Equation (\ref{eq:E3})) connects
the two guides through one coupling quantity only, the shunt admittance
$Y_{s}$. The dispersion Equation  (\ref{JK4-1} or \ref{eq:JKP})
reduces to Equation (\ref{JKb}), with $E_{i}=Y_{s}B_{i}$. This equation
is obtained for the choice of \textit{specific} admittances and impedances,
corresponding to the choice of acoustic pressure and velocity (of
the planar mode) as basic quantities for the considered 4-ports. This
choice is convenient for the description of the perforation effects,
but when the series impedance $Z_{a}$ is ignored, it is easier to
use flow rates instead of velocities (therefore to use \textit{acoustic}
admittances and impedances). For this purpose the impedances and admittances
need to be modified, and Equation  (\ref{JKb}) becomes: 
\begin{multline}
(\cosh\Gamma-A_{1}-\frac{1}{2}\bar{Y}_{p}\overline{B}_{1})(\cosh\Gamma-A_{2}-\frac{1}{2}\bar{Y}_{p}\overline{B}_{2})\\
=\frac{1}{4}\bar{Y}_{p}^{2}\overline{B}_{1}\overline{B}_{2},\label{eq:JJKP}
\end{multline}
where $\overline{B}_{1}=B_{1}/S_{1}$ $\ \overline{B}_{2}=B_{2}/S_{2}$
and the \textit{acoustic} admittance $\bar{Y}_{p}$ is given by:
\begin{equation}
Y_{s}=\frac{1}{2}\left(\frac{1}{S_{1}}+\frac{1}{S_{2}}\right)\bar{Y}_{p}\label{eq:YS}
\end{equation}
The bars above the symbols indicate \textit{acoustic} impedances or
admittances. When the radius of the perforation is very small, a simple
formula can be chosen: 
\begin{equation}
\frac{1}{\bar{Y}_{p}}=\frac{j\omega(\rho_{1}+\rho_{2})}{4r_{p}},\label{eq:E8}
\end{equation}
where $r_{p}$ is the radius of a circular perforation or the equivalent
radius when the perforation is not circular. Another form of the dispersion
equation is useful: 
\begin{equation}
\frac{2}{\bar{Y}_{p}}=\frac{\overline{B}_{1}}{\cosh\Gamma-A_{1}}+\frac{\overline{B}_{2}}{\cosh\Gamma-A_{2}}\ .\label{40}
\end{equation}
For the calculation of the eigenvectors, we make the choice of an
asymmetric cell, and use Equations (\ref{JK7}) and (\ref{JK8}).
For the eigenvalue $\lambda=\exp(\Gamma),$ it is found: 
\begin{equation}
\mathbf{W}_{L1}^{(i)}=\frac{v_{0}}{\gamma_{1}(\cosh\Gamma-A_{1})}\left(\begin{array}{c}
\overline{B}_{1}S_{1}e^{-\Gamma}\\
1-A_{1}e^{-\Gamma}
\end{array}\right),\label{200}
\end{equation}
and similarly for Guide $2$ (with a change in sign). Using Equation
 (\ref{40}), the pressure ratio is found to be: 
\begin{equation}
\frac{p_{1}}{p_{2}}=1+\frac{A_{2}-\cosh\Gamma}{\frac{1}{2}\bar{Y}_{p}\overline{B}_{2}}.\label{5}
\end{equation}


\subsection{Definition of the coupling coefficient\label{IIIA}}

The discriminant $\Delta$ of the quadratic equation in $\cosh\Gamma$
(Equation (\ref{eq:JJKP})) can be written by exhibiting a \emph{coupling
coefficient} $\mathcal{C}$, as follows: 
\begin{eqnarray}
\Delta & = & (A_{1}-A_{2})^{2}\left[1+2\frac{\overline{B}_{1}-\overline{B}_{2}}{\overline{B}_{1}+\overline{B}_{2}}\mathcal{C}+\mathcal{C}^{2}\,,\right]\label{55a}\\
 & \text{where} & \text{ \ }\mathcal{C}=\frac{1}{2}\bar{Y}_{p}\frac{\overline{B}_{1}+\overline{B}_{2}}{A_{1}-A_{2}}\text{ \ .}\label{4}
\end{eqnarray}
The coupling coefficient $\mathcal{C}$ is proportional to the perforation
admittance $\bar{Y}_{p}$, and inversely proportional to the difference
between the coefficients $A_{1}$ and $A_{2}$ of the two guides,
which are characteristic of the propagation into each guide separately.
For identical guides, $\mathcal{C}$ is infinite; at low frequencies,
the admittance is large, so is $\mathcal{C}$. Two extreme cases can
therefore be distinguished.

\subsubsection{The weak coupling limit}

When the coupling coefficient \ $\mathcal{C}$ is small, the following
solution is found:

\begin{gather}
\cosh\Gamma=A_{1}+\frac{1}{2}\bar{Y}_{p}\overline{B_{1}}+\mathcal{C}^{2}\frac{(A_{1}-A_{2})\overline{B}_{1}\overline{B}_{2}}{(\overline{B}_{1}+\overline{B}_{2})^{2}}+O(\mathcal{C}^{3})\label{JK20}\\
=A_{1}+\frac{1}{2}\bar{Y}_{p}\overline{B_{1}}+\frac{1}{4}\frac{\bar{Y}_{p}^{2}\overline{B}_{1}\overline{B}_{2}}{A_{1}-A_{2}}+O(\mathcal{C}^{3}).\notag\\
\frac{p_{2}}{p_{1}}=2\mathcal{C}^{2}\frac{(A_{2}-A_{1})B_{2}}{(\overline{B}_{1}+\overline{B}_{2})^{2}\bar{Y}_{p}}+O(\mathcal{C}^{3})\\
=-\mathcal{C}\frac{\overline{B}_{2}}{(\overline{B}_{1}+\overline{B}_{2})}+O(\mathcal{C}^{3}).\notag
\end{gather}
The mode $\Gamma^{\prime}$ is obtained by exchanging the subscripts
$1$ and $2.$ For a very weak coupling, the solution (\ref{JK20})
can be interpreted, at the first order of $\mathcal{C}$, as follows:
the medium 2 acts as an equivalent impedance $Z_{eq}=\bar{Y}_{p}^{-1}$
on the medium 1. The pressure becomes very small in the medium 2.

\subsubsection{The strong coupling limit; locally reacting impedance\label{sub:The-strong-coupling}}

Strong coupling occurs when the media are not very different or when
the perforation effect is strong (large opening of the perforation
and/or low frequencies). The solutions $\cosh\Gamma$ and $\cosh\Gamma^{\prime}$
of Equation  (\ref{eq:JJKP}) can be written as a series expansion
with respect to $\mathcal{C}^{-1}$: 
\begin{eqnarray}
\cosh\Gamma & = & \frac{A_{1}\overline{B}_{2}+A_{2}\overline{B}_{1}}{\overline{B}_{1}+\overline{B}_{2}}\label{4a}\\
 & - & \frac{1}{\mathcal{C}}\frac{\overline{B}_{1}\overline{B}_{2}}{\left(\overline{B}_{1}+\overline{B}_{2}\right)^{2}}(A_{1}-A_{2})\left[1+O(\frac{1}{\mathcal{C}})\right]\ ,\notag\\
\cosh\Gamma^{\prime} & = & \frac{1}{2}\bar{Y}_{p}(\overline{B}_{1}+\overline{B}_{2})+\frac{A_{1}\overline{B}_{1}+A_{2}\overline{B}_{2}}{\overline{B}_{1}+\overline{B}_{2}}\label{4b}\\
 &  & +\frac{1}{\mathcal{C}}\frac{\overline{B}_{1}\overline{B}_{2}}{\left(\overline{B}_{1}+\overline{B}_{2}\right)^{2}}(A_{1}-A_{2})\left[1+O(\frac{1}{\mathcal{C}})\right].\notag
\end{eqnarray}
The mode $\Gamma$ is an average value of the two propagation constants
given by $\cosh\Gamma=A_{1}$ and $\cosh\Gamma=A_{2}$. The mode $\Gamma^{\prime}$
is a generalization of the flute mode (see \cite{Kergomard1994}):
it is strongly evanescent at low frequencies and for large perforations
(large $\bar{Y}_{p}$ ). This mode was also given by Pierce \cite{Pierce1991}
for the continuous case of a perforated tube muffler.\ The pressure
ratios corresponding to the two modes $\Gamma$ and $\Gamma^{\prime}$
are: 
\begin{eqnarray}
\left[\frac{p_{1}}{p_{2}}\right] & = & 1+\frac{1}{\mathcal{C}}+\frac{1}{\mathcal{C}^{2}}\frac{\overline{B}_{1}}{\overline{B}_{1}+\overline{B}_{2}}\left[1+O(\frac{1}{\mathcal{C}})\right]\label{6}\\
\left[\frac{p_{1}}{p_{2}}\right]^{\prime} & = & -\frac{\overline{B}_{1}+\overline{B}_{2}}{\overline{B}_{2}}+\frac{1}{\mathcal{C}}\frac{\overline{B}_{1}}{\overline{B}_{2}}-\frac{1}{\mathcal{C}^{2}}\frac{\overline{B}_{1}}{\overline{B}_{1}+\overline{B}_{2}}\left[1+O(\frac{1}{\mathcal{C}})\right]\notag.\label{7}
\end{eqnarray}
The first solution corresponds to a modified planar mode while for
the second, the two pressures are opposite in phase, at least at low
frequencies. For the case where the mode $\Gamma^{\prime}$ is strongly
evanescent, only the first mode can be taken into account, and the
effect of the second medium on the first one (or vice-versa) can be
represented by an equivalent impedance $Z_{eq}$, the second one being
equivalent to a locally reacting medium. This impedance can be found
from the following dispersion equation: 
\begin{equation}
\cosh\Gamma=A_{1}+\frac{1}{2}\bar{Z}_{eq}^{-1}\overline{B}_{1}\text{ ,}\label{8}
\end{equation}
obtained by searching for the eigenvalue of the periodic medium built
with a cell described by the following transfer matrix: 
\[
\left(\begin{array}{cc}
A_{1} & \overline{B}_{1}\\
\overline{C}_{1} & A_{1}
\end{array}\right)\left(\begin{array}{cc}
1 & 0\\
\bar{Z}_{eq}^{-1} & 1
\end{array}\right).
\]
From Equations (\ref{4a},\ref{8}), the following equivalent impedance
$Z_{eq}$ is found to be: 
\begin{equation}
\bar{Z}_{eq}=\frac{\overline{B}_{2}}{\bar{Y}_{p}(\overline{B}_{1}+\overline{B}_{2})}\left[1+O(\frac{1}{\mathcal{C}})\right]+\frac{\overline{B}_{1}+\overline{B}_{2}}{2(A_{2}-A_{1})}.\label{9}
\end{equation}
Obviously this concept of equivalent impedance is especially interesting
if it does not depend on the first medium: this situation occurs if
$\ \ \overline{B}_{1}<<\overline{B}_{2}$, and $A_{1}<<A_{2}$. As
expected, it can be checked that these conditions of local reaction
are fulfilled in particular when the cells of the second medium are
uncoupled, e.g. thanks to closed walls between them. If these conditions
are satisfied the equivalent impedance is the sum of the perforation
impedance and the half of the input impedance of a cell of the second
medium, closed by a rigid wall: 
\begin{equation}
\bar{Z}_{eq}=\frac{1}{\bar{Y}_{p}}+\frac{1}{2}\frac{\overline{B}_{2}}{A_{2}-1},\label{10}
\end{equation}

and corresponds to Helmholtz resonators branched on Guide 1. 

\section{Application examples of lattices with finite length\label{sec:Applications}}
\subsection{Definition of the geometries considered}

\begin{table}
\begin{centering}
\begin{tabular}{|c|c|}
\hline 
Nb of cells $n$  & 5 \tabularnewline
\hline 
Cell length $(2\ell)$ $(m)$  & $2.17\,10^{-1}$ \tabularnewline
\hline 
Cross section $S{}_{1}\,(m^{2})$  & $3.14\,10^{-2}$ \tabularnewline
\hline 
Cross section $S_{2}\,(m^{2})$  & $3.46\,10^{-2}$ \tabularnewline
\hline 
Perf. radius $r_{p}\,(m)$ & $3.9\,10^{-2}$ \tabularnewline
\hline 
Perf. open area ratio $\sigma_{p}$  & $3.5\,10^{-2}$ \tabularnewline
\hline 
Diaph. radius $r_{d}\,(m)$ (Fig.\ref{fig:05}) & $0.105\:\textrm{and}\:0.104$\tabularnewline
\hline 
Diaph. open area ratio $\sigma_{d}$ (Fig.\ref{fig:05}) & $1\,\textrm{and}\,0.98$ \tabularnewline
\hline 
Diaph. radius $r_{d}\,(m)$ (Fig.\ref{fig:06}) & $0.018\,\textrm{and}\,0$\tabularnewline
\hline 
Diaph. open area ratio $\sigma_{d}$ (Fig.\ref{fig:06}) & $0.03\,\textrm{and}\,0$ \tabularnewline
\hline 
\end{tabular}
\par\end{centering}

\caption{Geometrical parameters of the finite lattice \label{tab:GEOM}}
\end{table}

We consider the geometries of finite length lattices shown in Figure
\ref{fig:Three-Cases}. In particular we focus on parameters (see
Table \ref{tab:GEOM}) that correspond to the strong coupling case
(see Section \ref{sub:The-strong-coupling}). Therefore
Guide 2 is strongly coupled to Guide 1 and acts as an acoustic wall
treatment on Guide 1. 

If the series impedance of the perforation is ignored, the perforation
matrix (Equation (\ref{eq:E3})) is completely defined by choosing
the specific shunt admittance $Y_{s}$. A simple formula, sufficient
for our purpose is chosen for the acoustic admittance $\bar{Y}_{p}$,
according to Equation (\ref{eq:E8}) as follows:

\begin{equation}
\frac{1}{\bar{Y_{p}}}=R+j\omega\rho/2r_{p},\label{eq:YP}
\end{equation}

where $R=2\sqrt{2\eta\rho\omega}$ is a small resistive term \cite{Ingard1953}
describing viscous losses near the perforation and $\eta$ is the
shear viscosity of the fluid. This allows limiting the resonance heights.
Equation (\ref{eq:YP}) is valid for small perforations, and also
for a length between perforations sufficiently large in comparison
with the transverse dimensions of the guide. This issue, which is
important for the understanding of the transition between discrete
and continuous descriptions, is discussed in \cite{Kergomard1994}.
For the sake of simplicity we consider here that an equivalent radius
exists (see e.g. \cite{rschevkin1963}). We define the  radius $r_p$ of the perforation  thanks to its open area ratio in one cell of Guide 1 $\sigma_{p}=\pi r_p^2 / (2\ell a_1) $. The shunt admittance $\bar{Y}_{p}$
vanishes when the perforation is closed.

In air, the transfer matrix for the planar mode along one uncoupled
portion of length $\ell$ is: 
\begin{equation}
\mathbf{T}=\left(\begin{array}{cc}
\cos(k\ell) & jZ_{c}\sin(k\ell)\,\\
j\sin(k\ell)/Z_{c} & \cos(k\ell)
\end{array}\right)\,.
\end{equation}
If propagation losses are ignored, $Z_{c}=\rho c$ and $k=\omega/c$
are the characteristic impedance and wavenumber of the medium (air)
filling the guides 1 and 2. For Guide 1, the transfer matrix along
length $\ell$ is: $\mathbf{T}_{1R}=\mathbf{T}_{1L}=\mathbf{T}$,
and $\mathbf{T}_{1}=\mathbf{T}^{2}$ for the length $2\ell$ (Equation
(\ref{eq:E6})). Inside Guide 2, we write for the length $\ell$ on the
left of a perforation $\mathbf{T}_{2L}=\mathbf{D_{d}\,T}$, and $\mathbf{T}_{2R}=\mathbf{T\,D_{d}}$ on the right. Therefore 
 $\mathbf{T}_{2}\mathbf{=\mathbf{TD_{d}^{2}}\,T}$
for the length $2\ell$ . The matrix 
\begin{equation}
\mathbf{D_{d}}=\left(\begin{array}{cc}
1 & \bar{Z}_{d}/2\\
0 & 1
\end{array}\right)
\end{equation}
corresponds to the presence of a diaphragm within Guide 2, and introduces
the non-homogeneity between the two guides. For the sake of simplicity, the (\emph{acoustic})
impedance is $\bar{Z}_{d}=j\omega\rho S_{2}/2r_{d}\,\left(1-\sqrt{\sigma_{d}}\right)$,
where $r_{d}$ is the opening radius of a diaphragm without thickness, and $\sigma_{d}=(r_{d}/a_{2})^{2}$ is its open area ratio.
This is a crude simplification of Fock's formula \cite{Fock}.

\subsection{From  homogeneous to  slightly non-homogeneous lattices}

\begin{figure}[t]
\centering{}\includegraphics[width=8.5cm]{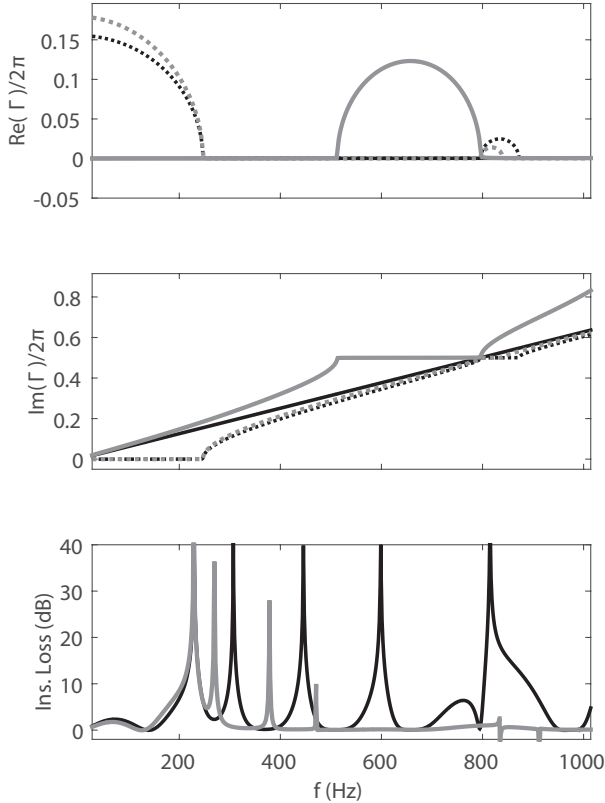}\caption{Dispersion curves $\Gamma(\omega)$ (top and center) and Insertion
loss $IL(\omega)$ (bottom). Black lines: homogeneous lattice ($\sigma_{d}=1$).
Grey lines: slightly non-homogeneous lattice ($\sigma_{d}=0.98$). Solid lines: mode with propagation constant $\Gamma$. Dotted lines: mode with propagation constant $\Gamma'$ (notice that $\Gamma'$ is mildly affected by the non-homogeneity). \label{fig:05}}
\end{figure}

Consider the homogeneous lattice  shown in Figure \ref{fig:Three-Cases}c,
where the same medium fills the waveguides 1 and 2. No diaphragm is
placed in Waveguide 2. The lattice is said to be homogeneous. The
wavenumber and characteristic impedance are identical for the two
waveguides, and $A_{1}=A_{2}=A=\cos(2k\ell)$. Equation (\ref{40})
gives the planar mode with constant $\Gamma$ and the flute mode with constant $\Gamma'$: 
\begin{align}
\begin{cases}
\cosh\Gamma & =\cos(2k\ell)\,,\\
\cosh\Gamma^{\prime} & =\cos(2k\ell)+j\bar{Y}_{p}Z_{c}\left[\frac{1}{S_{1}}+\frac{1}{S_{2}}\right]\sin(2k\ell).
\end{cases}\,\label{eq:HOM}
\end{align}
They correspond to the first term of Equations   (\ref{4a}) and (\ref{4b}),
respectively (the coupling coefficient $\mathcal{C}$ is infinite). Their variation with frequency is shown in Figure \ref{fig:05} (top and center).
The planar mode, with $\Gamma=2jk\ell$, is not dispersive and is unaffected by the perforation admittance
$\bar{Y}_{p}$. The flute mode, with constant $\Gamma^{\prime}$, is evanescent at low frequencies, because $\bar{Y}_{p}$
is high, yielding $\cosh\Gamma^{\prime}>1$ (or $Re(\Gamma)>0$ and
$Im(\Gamma)=0$). The mode $\Gamma^{\prime}$ is cut on at $f=247$
Hz. Above this frequency the two modes propagate within the lattice. 

\begin{table}[tb]
\begin{centering}
\begin{tabular}{|c|cc|}
\hline 
  & \multicolumn{2}{c|}{Lattice mode}\tabularnewline
\hline 
Freq.(Hz)  & $\Gamma$  & $\Gamma'$ \tabularnewline
\hline 
247{*} & - & Eq.\ref{JK26}\tabularnewline
\hline 
509 & Eq.\ref{JK28} & -\tabularnewline
\hline 
797{*} & Eq.\ref{JK29} & Eq.\ref{JK29}\tabularnewline
\hline 
839 & - & Eq.\ref{JK28}\tabularnewline
\hline 
\end{tabular}
\par\end{centering}

\caption{\label{tab:Cut-offA}Cut-off frequencies of the slighty non-homogenous
modes in Figure \ref{fig:05} (in grey). Symbol {*} denotes a cut-off
frequency that does not depend on the diaphragm radius $r_{d}$.}
\end{table}

Looking at frequencies below 797 Hz (this limit is explained later
on), the Insertion Loss obtained for the homogeneous lattice (Figure
\ref{fig:05} (bottom) is similar to the results presented
in \cite{Sullivan1978} (Figure 12). The Insertion Loss curve exhibits
a low frequency behaviour similar to that of an expansion chamber
(driven by the expansion ratio $S_{1}/S_{2}$). At higher frequencies,
high Insertion Loss peaks, limited by losses, are observed.

As a first result of the present analysis, this behaviour change in
Insertion Loss can be associated to the number of propagating modes
(here 1 or 2) within the lattice. A second result is that the frequencies
at which the Insertion Loss is maximum are given by Equation (\ref{eq:E70}).
It can be shown that the maximum for Insertion Loss is resonant (limited by losses) only
if the two terms of Equation (\ref{eq:E70}) have the same order of
magnitude and if the lattice length is finite.

Above 797 Hz, a stop band of Bragg type appears for the flute mode
$\Gamma^{\prime}$, with a high Insertion Loss (Figure \ref{fig:05},
bottom). In a Bragg stop band (or Bragg resonance due to spatial periodicity)
we have $Im(\Gamma)$ is a constant $n\pi$, where $n$ is an integer and
$Re(\Gamma)$ is positive but remains finite \cite{Elachi1976}. Unlike Figure
\ref{fig:05}, this behaviour is not reported in \cite{Sullivan1978},
presumably because the spatial periodicity of the lattice that \cite{Sullivan1978}
used makes the Bragg stop band out of the frequency band presented
(unfortunately this periodicity is not mentioned in \cite{Sullivan1978}).

Let us now consider a slightly non-homogeneous lattice. The diaphragms are now slightly closed ($\sigma_{d}=(r_{d}/a_{2})^{2}\simeq0.98)$. The solution
of the dispersion Equation (\ref{40}) for the  mode $\Gamma$ of the
non-homogenous lattice is: 
\begin{equation}
\cosh\Gamma=\frac{1}{2}\left(A_{1}+A_{2}+\frac{1}{2}\bar{Y}_{p}\left[\overline{B}_{1}+\overline{B}_{2}\right]-\sqrt{\Delta}\right)\,,\label{eq:PlaneMode}
\end{equation}
where $\Delta$ is given by Equation (\ref{55a}). The second solution for
the mode $\Gamma'$ is given by the same result, changing the sign before $\Delta$.
 The mode $\Gamma'$ is not really affected by the added mass. In particular,
its first cut-off frequency remains unchanged at 247 Hz, like for the
homogeneous lattice (the cut-off frequencies for the slightly non-homogeneous
lattice are given in Table \ref{tab:Cut-offA}). This is explained
by symmetry properties discussed in Section \ref{dtw},
which imply that the velocity within the diaphragms vanishes at that
particular frequency.

However the mass added by the diaphragms strongly modifies the
mode $\Gamma$: the mode is now dispersive, with a phase velocity lower than
that the  planar mode of the homogeneous lattice (see Figure \ref{fig:05}, center). This slowdown of the  mode $\Gamma$ explains why the resonant peaks of the
Insertion Loss  are
shifted towards the low frequencies, compared to those of the homogeneous lattice.
A Bragg stop band also appears for the mode $\Gamma$ between
509 and 797 Hz (see Figure \ref{fig:05}, top): only the mode $\Gamma'$  propagates and the Insertion Loss curve  does not exhibit resonant peaks because the mode $\Gamma$ is evanescent. For the mode $\Gamma^{\prime}$, the width of the the Bragg stop band is reduced
to the interval $[797,839]\:Hz$, where the Insertion Loss vanishes.
This issue could be further investigated. 

In these simulations the number of cells of the lattice is limited to
$n=5$ for the sake of readability of the Insertion Loss curves. Indeed,
increasing $n$ increases the number of resonant peaks,
according to Equation (\ref{eq:E70}). But it can be checked the computation does not encounter  numerical difficulties even for high $n$ (and/or highly evanescent
modes), thanks to the impedance matrix formalism.

 Summarizing the effect of the non-homogeneity
induced by the diaphragms, the width of frequency bands where
two modes propagate is reduced, and therefore the possibility of resonant
Insertion Loss as well.

\subsection{From ducts with branched Helmholtz resonators to  strongly non-homogeneous
lattices}

\begin{figure}[t]
\centering{}\includegraphics[width=8.5cm]{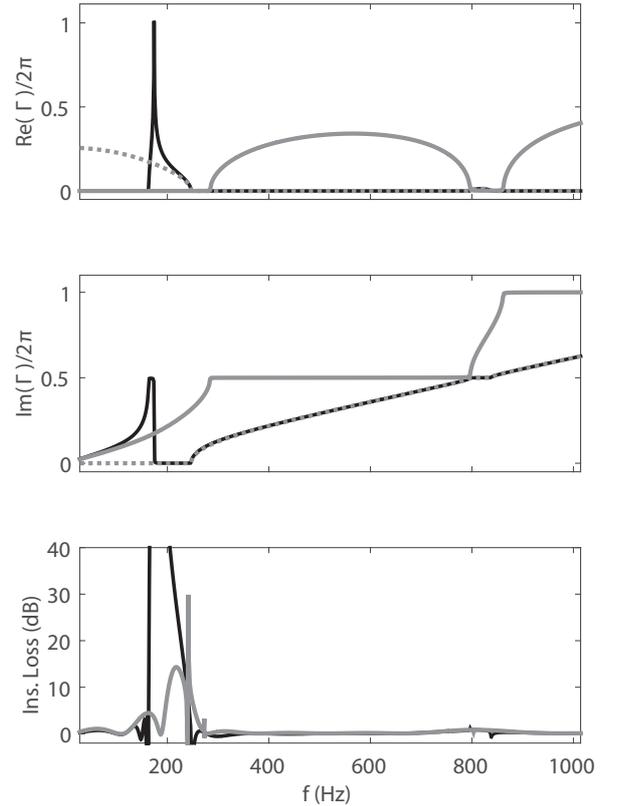}\caption{Dispersion curves $\Gamma(\omega)$ (top and center) and Insertion
loss $IL(\omega)$ (bottom). Black lines: branched Helmholtz resonators ($\sigma_{d}=0$).
Grey lines: strongly non-homogeneous lattice ($\sigma_{d}=0.03$). Solid lines: mode with constant $\Gamma$. Dotted lines: mode with constant $\Gamma'$ (notice that for non-coupled resonators, there is only one mode).\label{fig:06} }
\end{figure}

Let us now start from another classical muffler configuration: the branched
Helmholtz resonators (Figure \ref{fig:Three-Cases}b), which are locally
reacting. There is only one mode in the lattice. It is given by Equation
(\ref{8}): 
\begin{equation}
\cosh\Gamma^{h}=\cos(2k\ell)\,+j\frac{\rho c}{2S_{1}}\bar{Z}_{eq}^{-1}\sin(2k\ell)\label{eq:HELMH}
\end{equation}
where, according to Equation (\ref{10}), the input impedance of one
resonator is $\bar{Z}_{eq}=1/\bar{Y}_{p}-j(\rho c/2S_{2})\,\cot (k\ell)$.
This mode $\Gamma^{h}$ is strongly dispersive (see Figure \ref{fig:06}) and exhibits a stop band within the band $[f_{1}^{h},f_{2}^{h}]=[162,247]$
Hz, which can be called ``Helmholtz stop band''. This band is a resonance
stop band (indeed, the Helmholtz resonance frequency $f_{r}$ is given by $\bar{Z}_{eq}=0$). $Im(\Gamma)=m\pi$ below $f_{r}$ , and $Im(\Gamma)=m'\pi$
above $f_{r}$, $m$ being an even integer  and $m'$ an odd integer (or vice versa). Moreover, if there are no losses, $Re(\Gamma)$ is infinite at $f_{r}$.
The lower bound of the stop band  $f_{1}^{h}$, given by $\cosh\Gamma^{h}=-1$
or
\begin{equation}
\frac{2}{\bar{Y}_{p}}=jZ_{c}\left(-\frac{1}{S_{1}}\tan(k\ell)+\frac{1}{S_{2}}\cot(k\ell)\right)\:.
\end{equation}
The upper bound is  $f_{2}^{h}$, given by $\cosh\Gamma^{h}=1$
or 
\begin{equation}
\frac{2}{\bar{Y}_{p}}=jZ_{c}\left(\frac{1}{S_{1}}+\frac{1}{S_{2}}\right)\tan(k\ell)\,.\label{eq:CutOnHelmh}
\end{equation}

\begin{table}[tb]
\begin{centering}
\begin{tabular}{|c|cc|}
\hline 
  & \multicolumn{2}{c|}{Lattice Mode}\tabularnewline
\hline 
Freq.(Hz)  & $\Gamma$  & $\Gamma'$ \tabularnewline
\hline 
247{*} & - & Eq.\ref{JK26}\tabularnewline
\hline 
284 & Eq.\ref{JK28} & -\tabularnewline
\hline 
797{*} & Eq.\ref{JK29} & Eq.\ref{JK29}\tabularnewline
\hline 
839 & - & Eq.\ref{JK28}\tabularnewline
\hline 
861 & Eq.\ref{JK27} & \_\tabularnewline
\hline 
\end{tabular}
\par\end{centering}

\caption{\label{tab:Cut-offB}Cut-off frequencies of strongly non-homogeneous
modes in Figure \ref{fig:06} (in grey). Symbol {*} denotes a cut-off
frequency that does not depend on the diaphragm radius $r_{d}$.}
\end{table}

Notice that this cut-off frequency $f_{2}^{h}$ defined by Equation
(\ref{eq:CutOnHelmh}) is the the first cut-off frequency
of the mode $\Gamma^{\prime}$ defined by Equation (\ref{JK26}).
In particular, $f_{2}^{h}$ remains unchanged when a diaphragm is
open between resonators, because the velocity within the diaphragms
vanishes at this particular frequency (Section \ref{dtw}).

 Consider now Figure \ref{fig:06}, which shows the effect of a strong inhomogeneity of the lattice.  It appears that the
mode $\Gamma'$  ($\sigma_{d}=(r_{d}/a_{2})^{2}\approx0.03$),
is cut off exactly at the upper bound of the Helmholtz stop band $f_{2}^{h}$. The
cut-off frequencies for the strongly non-homogeneous lattice are given
in Table \ref{tab:Cut-offB}. The mode $\Gamma$  propagates at low frequencies
and is evanescent for $f$ lying within $[284,797]$ Hz. This implies
that even a small opening of the diaphragms between resonators entails
that the stop band of the branched Helmholtz resonators disappears.
In particular, the singularity of $Re(\Gamma^{h})$ at $f_{r}$,
which is a characteristic of the Helmholtz stop band, is lost.

Summarizing the effect of the coupling of the branched resonators,  the two  modes have very different behaviours  for all the frequencies considered. This tends
to limit the Insertion Loss of the strongly non-homogeneous lattice
compared to that of branched resonators.

\section{Conclusion }

The analytical approach proposed is able to describe a wide variety of periodically
coupled waveguides. For two classical examples
of applications (homogeneous lattices and branched Helmholtz resonators),
the model  shows how the frequency behaviour of the Insertion
Loss, can be  attributed either to  the properties of the
medium (dispersion within the lattice), or to the boundary conditions
and the finite length. Moreover, the introduction
of a non-homogeneity within the lattice, by means of an added mass
in one of the waveguides, illustrates how the properties (dispersion and Insertion
Loss) of the two classical examples are modified, and how this can be interpreted.

A coupling coefficient is useful for the study of  the transition between
local and non-local reaction of one waveguide to the other. In practice,
the model have shown that a very small coupling between (locally reacting)
Helmholtz resonators is sufficient to obtain a lattice where the local
reaction vanishes. A particular situation is encountered when an interaction between Bragg and Helmholtz 
stop bands occurs. How this could be combined with finite length
effects for sound attenuation purpose could be further investigated.

Other types of non-homogeneity, like the presence of dissipative media
(porous materials described as equivalent fluids) or varying cross sections
are in the scope of the method, provided that coupling of the evanescent
modes created by two singularities does not
occur, i.e., perforations are sufficiently spaced.

Arguments can be found for ignoring the series impedance
of the perforation, but this restricts applications to cases where the frequency is low and 
the perforation radius is small compared to the waveguides radii.
The knowledge of appropriated expressions for series impedance and
shunt admittance of the perforation would be  required for practical
application of the model to a particular geometry. An issue of interest
could be the effects of the series impedance $Z_{a}$
on the properties of a finite lattice at higher frequencies. Precise values of the perforation admittance and impedance remain a topic of further investigation, in particular when the frequency increases. This can be done either with numerical methods or with measurements.

Application can be done to different kind of devices, such a silencers or sample of 1D metamaterials of finite length.   To a certain extent, it could be possible to divide their design in two steps: first an optimization of the Insertion Loss with respect to given values of the perforation parameters, then a determination of the geometry corresponding to these parameters. 

With the same model, further investigation could be done on dissipation effects, either in the perforations or in the waveguides. Mean flow or nonlinear effects would require different models.

\appendix

\setcounter{equation}{0} 
\renewcommand{\theequation}{A\arabic{equation}}

\section*{Appendix A : Derivation of the eigenvalues and eigenvectors of the transfer
matrix\label{sec:Appendix:-Derivation-of}}
\renewcommand{\thesubsection}{A\arabic{subsection}}

\subsection{Eigenvalues, dispersion equation}

For the sake of simplicity, the eigenvalues $\lambda$ and eigenvectors,
denoted $\mathcal{W}_{L}$, are first sought for a generic, asymmetric
cell. They are solutions of the 4th-order equation:

\begin{equation}
(\mathcal{P}_{\mathcal{F}}\mathcal{T}\ -\lambda\mathcal{I\mathrm{)}W}_{L}\mathcal{=}\mathcal{O},\label{eq:E12}
\end{equation}
where $\mathcal{O}$ is the zero matrix of 4th-order. Using Equations
  (\ref{eq:E2}) and (\ref{eq:E6}), and a calculation based upon
sub-matrices, Equation (\ref{eq:E12}) can be rewritten as follows:
\begin{alignat}{1}
 & \begin{cases}
\left[\left(\gamma_{1}+\gamma_{2}\mathbf{M}\right)\mathbf{T}_{1}-\lambda\mathbf{I}\right]\mathbf{W}_{L1}+\gamma_{2}(\mathbf{I-M)T}_{2}\mathbf{W}_{L2} & =\mathbf{0}\,,\\
\gamma_{1}(\mathbf{I-M)T}_{1}\mathbf{W}_{L1}+\left[\left(\gamma_{2}+\gamma_{1}\mathbf{M}\right)\mathbf{T}_{2}-\lambda\mathbf{I}\right]\mathbf{W}_{L2} & =\mathbf{0}
\end{cases}\label{JK5}\\
 & \text{ \ if \ \ \ \ }\mathcal{W}_{L}\overset{def}{=}\,^{t}\left(\begin{array}{cc}
\mathbf{W}_{L1} & \mathbf{W}_{L2}\end{array}\right).\label{JK6}
\end{alignat}
Subtracting the two Equations (\ref{JK5}) leads to a new equation:
\begin{equation}
(\mathbf{MT}_{1}-\lambda\mathbf{I})\mathbf{W}_{L1}=(\mathbf{MT}_{2}-\lambda\mathbf{I})\mathbf{W}_{L2}.\label{eq:A4-1}
\end{equation}
Then, multiplying in System (\ref{JK5}) the first equation by $\gamma_{1}$
and the second equation by $\gamma_{2},$ and adding the resulting
equations, the following equation is obtained: 
\begin{equation}
\gamma_{1}(\mathbf{T}_{1}-\lambda\mathbf{I})\mathbf{W}_{L1}=-\gamma_{2}(\mathbf{T}_{2}-\lambda\mathbf{I})\mathbf{W}_{L2}\overset{def}{=}\mathbf{W}_{0}.\label{eq:A5-1}
\end{equation}
Then, writing for $i=1$ and $2$: 
\[
\mathbf{MT}_{i}-\lambda\mathbf{I=M(T}_{i}-\lambda\mathbf{I)+\lambda(M-I)}\,,
\]
and substituting in Equation (\ref{eq:A4-1}) the values of $\mathbf{W}_{0}$
given by Equation (\ref{eq:A5-1}), Equation (\ref{eq:A4-1}) can
be written as follows: 
\begin{multline}
(\mathbf{M}-\mathbf{I})+\mathbf{I}+\\
\mathbf{\lambda(M-I)}\left[\gamma_{2}(\mathbf{T}_{1}-\lambda\mathbf{I})^{-1}+\gamma_{1}(\mathbf{T}_{2}-\lambda\mathbf{I})^{-1}\right]\mathbf{W}_{0}=0.\label{JK3}
\end{multline}
Finally, multiplying Equation  (\ref{JK3}) by the matrix $2(\mathbf{M}-\mathbf{I)}^{-1}=\mathbf{K}-\mathbf{I,}$
where $\mathbf{K}$\ is given by Equation  (\ref{eq:E3}), it is
found that the 4th-order Equation  (\ref{eq:E12}) is equivalent to
the following 2nd-order equation: 
\begin{gather}
\mathbf{T}_{0}\mathbf{W}_{0}=\mathbf{0}\textrm{ where}\text{ }\label{360}\\
\mathbf{\mathbf{T}_{0}}\mathbf{=I}+\mathbf{K+}2\lambda\left[\gamma_{2}(\mathbf{T}_{1}-\lambda\mathbf{I})^{-1}+\gamma_{1}(\mathbf{T}_{2}-\lambda\mathbf{I})^{-1}\right]\label{eq:E14}
\end{gather}
Consequently, each eigenvalue $\lambda^{(i)}\ (i=1..4)$ is solution
of the general dispersion equation: 
\begin{equation}
\det\mathbf{T}_{0}=A_{0}D_{0}-B_{0}C_{0}=0.\label{JK4}
\end{equation}
Here the coefficients of the matrix $\mathbf{T}_{0}$ are denoted
$A_{0},$ $B_{0},$ $C_{0},$ $D_{0}$. Equation  (\ref{eq:E14})
gives their expression, which depends on the eigenvalue $\lambda$:
\begin{eqnarray}
A_{0} & = & 1+2\lambda\left[\frac{\gamma_{2}(D_{1}-\lambda)}{\Delta_{1}}+\frac{\gamma_{1}(D_{2}-\lambda)}{\Delta_{2}}\right],\label{eq:E17}\\
D_{0} & = & 1+2\lambda\left[\frac{\gamma_{2}(A_{1}-\lambda)}{\Delta_{1}}+\frac{\gamma_{1}(A_{2}-\lambda)}{\Delta_{2}}\right],\label{eq:E18}\\
B_{0} & = & \frac{1}{Y_{s}}-2\lambda\left[\gamma_{2}\frac{B_{1}}{\Delta_{1}}+\gamma_{1}\frac{B_{2}}{\Delta_{2}}\right],\label{eq:E19}\\
C_{0} & = & \frac{1}{Z_{a}}-2\lambda\left[\gamma_{2}\frac{C_{1}}{\Delta_{1}}+\gamma_{1}\frac{C_{2}}{\Delta_{2}}\right]\,,\label{eq:E20}
\end{eqnarray}
where the coefficients $A_{1,2}\,,B_{1,2}\,,C_{1,2}\,,D_{1,2}$ of
matrices $\mathbf{T}_{1,2}$ correspond to the diagonal blocks of
the 4th-order transfer matrix $\mathcal{T}$ (see Equations  (\ref{eq:E5})
and (\ref{eq:E6}), and where 
\begin{equation}
\Delta_{1,2}=\det(\mathbf{T}_{1,2}-\lambda\mathbf{I)=\lambda}^{2}-\lambda(A_{1,2}+D_{1,2})+\det\mathbf{T}_{1,2}.
\end{equation}
Equation  (\ref{JK4}) is a simplification of the dispersion equation given in \cite{Kergomard1994} (see Equation of this reference).  In the same reference, the
expression of the 4th-order equation for the unknown $\lambda$ is
given (see Equation (43) of this reference).

\subsection{Eigenvectors for an asymmetric cell \label{sub:Eigenvectors-an-asymetrical}}

The eigenvectors of the matrix $\mathcal{P}_{\mathcal{F}}\mathcal{T}$
(see Equation (\ref{E11})) can be obtained thanks to Equation (\ref{eq:A5-1})
by determining the 2nd-order vector $\mathbf{W_{0}}$ (Equation  (\ref{360})).
For each eigenvalue, this vector is defined apart from a constant
multiplicative value. The following general form is chosen: 
\begin{equation}
\mathbf{W}_{0}^{(i)}=v_{0}\left(\begin{array}{c}
w_{0}^{(i)}\\
-1
\end{array}\right),\ (i=1..4)\label{eq:E21}
\end{equation}
where $v_{0}$ is an arbitrary constant having the dimension of a
velocity, and $w_{0}^{(i)}$ are impedances associated to eigenvalues
$\lambda^{(i)}$. By construction, Expression (\ref{eq:E21}) fulfills
Equation (\ref{360}), which means that for each eigenvalues $\lambda^{(i)}$
we have:

\begin{equation}
w_{0}^{(i)}=\frac{B_{0}^{(i)}}{A_{0}^{(i)}}=\frac{D_{0}^{(i)}}{C_{0}^{(i)}}\ (i=1..4)).\label{eq:E22}
\end{equation}
Introducing the general form of $\mathbf{W_{0}}$ (Equation (\ref{eq:E21}))
into Expression (\ref{eq:A5-1}) gives the eigenvectors of an asymmetric
cell of the lattice. For the upper and lower rows, each column $i$
$(i=1..4)$ of the matrix $\mathcal{E_{L}}$ (Equation (\ref{E11}))
can be written as:

\begin{eqnarray}
\mathbf{W}_{L1}^{(i)} & = & \frac{v_{0}}{\gamma_{1}\Delta_{1}}\left(\begin{array}{c}
w_{0}^{(i)}\left[D_{1}-\lambda^{(i)}\right]+B_{1}\\
-w_{0}^{(i)}C_{1}-\left[A_{1}-\lambda^{(i)}\right]
\end{array}\right)\,,\label{JK7}\\
\mathbf{W}_{L2}^{(i)} & = & -\frac{v_{0}}{\gamma_{2}\Delta_{2}}\left(\begin{array}{c}
w_{0}^{(i)}\left[D_{2}-\lambda^{(i)}\right]+B_{2}\\
-w_{0}^{(i)}C_{2}-\left[A_{2}-\lambda^{(i)}\right]
\end{array}\right)\,.\label{JK8}
\end{eqnarray}
Finally each column $i$ $(i=1..4)$ of the matrix $\mathcal{E_{L}}$
is obtained by assembling the 2nd-order vectors $\mathbf{W}_{L1}^{(i)}$
and $\mathbf{W}_{L2}^{(i)}$ by Equation (\ref{JK6}).

\subsection{Eigenvectors for a symmetric cell}

The dispersion equation (\ref{JK4}) and the expression (\ref{JK7},\ref{JK8})) of the eigenvectors are general. A more useful expression
can be obtained if the propagation matrix $\mathcal{T}$ is spltted 
into two matrices in order to get a symmetric cell (see Section \ref{asc}
and Figure \ref{fig:Transfer-Matrix-symmetrical-and--1}). Consequently
$A_{0}=D_{0}$.

The transfer matrix of a symmetric cell (Equation (\ref{eq:E26}))
is written in the diagonal form (Equation (\ref{eq:E9})):comparing
this equation and Equation (\ref{eq:E9}), the columns of the eigenvector
matrix $\mathcal{E}$ (Equation (\ref{E11})) can be obtained from
the eigenvectors of the anti-symmetric cell $\mathcal{W}_{L}^{(i)}$
(Equations  (\ref{JK7},\ref{JK8})) by: 
\begin{equation}
\mathcal{W}^{(i)}=\left(\begin{array}{c}
\mathbf{W}_{1}^{(i)}\\
\mathbf{W}_{2}^{(i)}
\end{array}\right)=\mathcal{T}_{L}\mathcal{W}_{L}^{(i)}.\text{ }\label{eq:E29}
\end{equation}
Therefore, using Equation (\ref{eq:A5-1}), $\mathbf{W}_{1}^{(i)}$
is given by: 
\begin{eqnarray}
\mathbf{W}_{1}^{(i)} & = & \frac{1}{\gamma_{1}}(\mathbf{T}_{R1}-\lambda\mathbf{T}_{L1}^{-1})^{-1}\mathbf{W}_{0}^{(i)}\text{ \ or}\label{E30}\\
\mathbf{W}_{1}^{(i)} & = & \frac{v_{0}}{\Delta_{1}\gamma_{1}}\left(\begin{array}{c}
w_{0}^{(i)}D_{R1}\left(\delta_{1}-\lambda^{(i)}\right)+B_{R1}(\delta_{1}+\lambda^{(i)})\\
-w_{0}^{(i)}C_{R1}(\delta_{1}+\lambda^{(i)})-A_{R1}(\delta_{1}-\lambda^{(i)})
\end{array}\right)\notag\label{E31}
\end{eqnarray}
with $\delta_{1}=\det(\mathbf{T}_{L1})=\det(\mathbf{T}_{R1}).$ A
similar expression holds for the guide 2, with a sign $-$ before
$v_{0}.$ When the determinants are unity, the eigenvectors are given
by Equation (\ref{356}). 
\setcounter{equation}{0} 
\renewcommand{\theequation}{B\arabic{equation}}

\section*{Appendix B: Transfer matrix of a lattice of $n$ cells\label{sec:Appendix:-Transfer-matrix}}
\setcounter{subsection}{0} 
\renewcommand{\thesubsection}{B\arabic{subsection}}

\subsection{First form of the transfer matrix}

In order to simplify the calculation of the invert matrix of $\mathcal{E}$
(Equation (\ref{356})), it is convenient to write its first up-left
quarter in the form of a matrix product: 
\begin{equation}
\left(\begin{array}{cc}
\mathbf{W}_{1}^{(1)} & \mathbf{W}_{1}^{(2)}\end{array}\right)=v_{0}\mathbf{H}_{1}\mathbf{FD}^{-1/2}\label{eq:E36}
\end{equation}
where $\ \mathbf{H}_{1}=\left(\begin{array}{cc}
z_{1} & 0\\
0 & h_{1}
\end{array}\right),$ $\mathbf{F}=\left(\begin{array}{cc}
1 & 1\\
1 & -1
\end{array}\right)$ and \\
$\mathbf{D}=\left(\begin{array}{cc}
e^{\Gamma} & 0\\
0 & e^{-\Gamma}
\end{array}\right)$. Using similar notations, the three other quarters of
the matrix $\mathcal{E}$ are:

\begin{eqnarray}
\left(\begin{array}{cc}
\mathbf{W}_{1}^{(3)} & \mathbf{W}_{1}^{(4)}\end{array}\right) & = & v_{0}\mathbf{H}_{1}^{\prime}\mathbf{FD}^{\prime-1/2},\label{eq:E39}\\
\left(\begin{array}{cc}
\mathbf{W}_{2}^{(1)} & \mathbf{W}_{2}^{(2)}\end{array}\right) & = & -v_{0}\mathbf{H}_{2}\mathbf{FD}^{-1/2},\label{eq:E40}\\
\left(\begin{array}{cc}
\mathbf{W}_{2}^{(3)} & \mathbf{W}_{2}^{(4)}\end{array}\right) & = & -v_{0}\mathbf{H}_{2}^{\prime}\mathbf{FD}^{\prime-1/2}\text{ \ .}\label{eq:E41}
\end{eqnarray}
By assembling expressions (\ref{eq:E36}-\ref{eq:E41}), the eigenvector
matrix $\mathcal{E}=\left(\begin{array}{cccc}
\mathcal{W}^{(1)} & \mathcal{W}^{(2)} & \mathcal{W}^{(3)} & \mathcal{W}^{(4)}\end{array}\right)$ is written as:

\begin{equation}
\mathcal{E=}v_{0}\left(\begin{array}{cc}
\mathbf{\tilde{H}}_{1} & \mathbf{\tilde{H}}_{1}^{\prime}\\
-\mathbf{\tilde{H}}_{2} & -\mathbf{\tilde{H}}_{2}^{\prime}
\end{array}\right)\left(\begin{array}{cc}
\mathbf{F} & \mathbf{0}\\
\mathbf{0} & \mathbf{F}
\end{array}\right)\left(\begin{array}{cc}
\mathbf{D} & \mathbf{0}\\
\mathbf{0} & \mathbf{D}^{\prime}
\end{array}\right)^{-1/2}.\label{eq:E42}
\end{equation}
Thanks to this particular form for $\mathcal{E}$, that results from
reciprocity, the 4th-order transfer matrix for a lattice of $n$ symmetric
cells $(\mathcal{T}_{L}\mathcal{P}_{F}\mathcal{T}_{R})^{n}=\mathcal{\mathrm{(}E}\mathcal{D}^{n}\mathcal{E}^{-1})$
is obtained as:

\begin{eqnarray}
(\mathcal{T}_{L}\mathcal{P}_{F}\mathcal{T}_{R})^{n}= & \mathcal{E}\mathcal{D}^{n}\mathcal{E}^{-1}= & \left(\begin{array}{cc}
\mathbf{H}_{1} & \mathbf{H}_{1}^{\prime}\\
-\mathbf{H}_{2} & -\mathbf{H}_{2}^{\prime}
\end{array}\right)\left(\begin{array}{cc}
\mathbf{F} & \mathbf{0}\\
\mathbf{0} & \mathbf{F}
\end{array}\right)\notag\\
\times & \left(\begin{array}{cc}
\mathbf{D} & \mathbf{0}\\
\mathbf{0} & \mathbf{D}^{\prime}
\end{array}\right)^{n} & \begin{array}[t]{c}
\left(\begin{array}{cc}
\mathbf{F} & \mathbf{0}\\
\mathbf{0} & \mathbf{F}
\end{array}\right)^{-1}\left(\begin{array}{cc}
\mathbf{H}_{1} & \mathbf{H}_{1}^{\prime}\\
-\mathbf{H}_{2} & -\mathbf{H}_{2}^{\prime}
\end{array}\right)^{-1}\end{array}\label{eq:E43-1}
\end{eqnarray}

\subsection{Second form of the transfer matrix \label{sub:Characteristic-impedances-of}}

In order to derive the impedance matrix, a second form of the transfer
matrix is useful. The vector $\mathcal{V}$ (Equation (\ref{eq:E1}))
is replaced by a vector $\widetilde{\mathcal{V}}$ defined as follows:
\begin{equation}
\ \widetilde{\mathcal{V}}=\left(\begin{array}{c}
\mathbf{P}\\
\mathbf{U}
\end{array}\right)\text{ \ where }\mathbf{P}=\left(\begin{array}{c}
p_{1}\\
p_{2}
\end{array}\right)\text{ and }\mathbf{U}=\left(\begin{array}{c}
u_{1}\\
u_{2}
\end{array}\right),
\end{equation}
where $u_{i}=S_{i}v_{i}$ ($i=1,2)$ are the flow rates. Considering
the eigenvector matrix $\mathcal{E}$ (Equation (\ref{356})), a permutation
of the second and third rows and columns is required, as well as a
permutation of the second and third eigenvalues (see e.g. \cite{Strang2014}).
The following result is obtained:
\begin{gather}
\widetilde{\mathcal{E}}\mathcal{=}v_{0}\left(\begin{array}{cccc}
z_{1} & z_{1}^{\prime} & z_{1} & z_{1}^{\prime}\\
-z_{2} & -z_{2}^{\prime} & -z_{2} & -z_{2}^{\prime}\\
g_{1} & g_{1}^{\prime} & -g_{1} & -g_{1}^{\prime}\\
-g_{2} & -g_{2}^{\prime} & g_{2} & g_{2}^{\prime}
\end{array}\right)\widetilde{\mathcal{D}}^{-1/2};\nonumber \\
\nonumber \\
\widetilde{\mathcal{D}}=\left(\begin{array}{cccc}
e^{\Gamma} & 0 & 0 & 0\\
0 & e^{\Gamma^{\prime}} & 0 & 0\\
0 & 0 & e^{-\Gamma} & 0\\
0 & 0 & 0 & e^{-\Gamma^{\prime}}
\end{array}\right)\nonumber \\
\end{gather}
if $g_{i}=S_{i}h_{i}$, or equivalently:

\begin{eqnarray}
\widetilde{\mathcal{E}} & = & v_{0}\left(\begin{array}{cc}
\mathbf{Z} & \mathbf{0}\\
\mathbf{0} & \mathbf{G}
\end{array}\right)\left(\begin{array}{cc}
\mathbf{I} & \mathbf{I}\\
\mathbf{I} & -\mathbf{I}
\end{array}\right)\text{ \ with}\label{eq:E44}\\
\mathbf{Z} & = & \left(\begin{array}{cc}
z_{1} & z_{1}^{\prime}\\
-z_{2} & -z_{2}^{\prime}
\end{array}\right),\mathbf{G}=\left(\begin{array}{cc}
g_{1} & g_{1}^{\prime}\\
-g_{2} & -g_{2}^{\prime}
\end{array}\right).\label{eq:ZetG}
\end{eqnarray}
Finally the second from of the transfer matrix is:

\begin{equation}
\widetilde{\mathcal{V}}_{0}=\left(\begin{array}{cc}
\mathbf{Z} & \mathbf{0}\\
\mathbf{0} & \mathbf{G}
\end{array}\right)\left(\begin{array}{cc}
\mathbf{C_{n}} & \mathbf{S_{n}}\\
\mathbf{S_{n}} & \mathbf{C_{n}}
\end{array}\right)\left(\begin{array}{cc}
\mathbf{Z} & \mathbf{0}\\
\mathbf{0} & \mathbf{G}
\end{array}\right)^{-1}\widetilde{\mathcal{V}}_{n}\,,\label{450}
\end{equation}
with $\mathbf{C}_{n}=\left(\begin{array}{cc}
\cosh n\Gamma & 0\\
0 & \cosh n\Gamma^{\prime}
\end{array}\right),$ and $\mathbf{S}_{n}=\left(\begin{array}{cc}
\sinh n\Gamma & 0\\
0 & \sinh n\Gamma^{\prime}
\end{array}\right).$

\section*{Acknowledgments}
The authors thank Yves Aur\'egan for fruitful discussions during the preparation of the manuscript. 

\bibliographystyle{unsrt}

\end{document}